\documentclass[preprint,12pt,eqsecnum]{elsarticle}

\usepackage{graphicx}
\usepackage{dcolumn}
\usepackage{bm}
\usepackage{amsfonts}
\usepackage{latexsym}
\usepackage{amssymb}
\usepackage{verbatim}
\usepackage{amsthm}
\usepackage{amsmath}

\journal{Annals of Physics}

\begin{document}

\begin{frontmatter}



\title{
  Extension of the input-output relation for a Michelson
  interferometer to arbitrary coherent-state light sources: \\
  {\it \large --- Gravitational-wave detector and weak-value
    amplification --- }
}


\author{
  Kouji Nakamura\footnote{E-mail address: kouji.nakamura@nao.ac.jp},
  and
  Masa-Katsu Fujimoto\footnote{E-mail address: fujimoto.masa-katsu@nao.ac.jp}
}

\address{
  Gravitational-Wave Project Office,
  Optical and Infrared Astronomy Division,
  National Astronomical Observatory of Japan,
  Mitaka, Tokyo 181-8588, Japan
}

\begin{abstract}
  An extension of the input-output relation for a conventional
  Michelson interferometric gravitational-wave detector is carried out
  to treat an arbitrary coherent state for the injected optical beam.
  This extension is one of necessary researches toward the
  clarification of the relation between conventional
  gravitational-wave detectors and a simple model of a
  gravitational-wave detector inspired by weak-measurements in
  [A.~Nishizawa, Phys. Rev. A {\bf 92} (2015), 032123.].
  The derived input-output relation describes not only a conventional
  Michelson-interferometric gravitational-wave detector but also the
  situation of weak measurements.
  As a result, we may say that a conventional Michelson
  gravitational-wave detector already includes the essence of the
  weak-value amplification as the reduction of the quantum noise from
  the light source through the measurement at the dark port.
\end{abstract}

\begin{keyword}
  Gravitational-wave detector, weak-value amplification


\end{keyword}

\end{frontmatter}


\section{Introduction}
\label{sec:introduction}


Weak measurements and their weak-value amplifications have been
currently discussed by many researchers since their proposal by
Aharonov, Albert, and Vaidman in
1988~\cite{Y.Aharonov-D.Z.Albert-L.Vaidman-1988}.
In particular, the weak-value amplification has been regarded as one
of techniques that has been used in a variety of experimental settings
to permit the precise measurement of small
parameters~\cite{N.W.M.Ritchie-J.G.Story-R.G.Hulet-1991-etc}.
This paper is motivated by these researches on the precise
measurements in quantum theory.


As well-known,  one of typical examples of precise measurements is the
gravitational-wave detection.
Recently, gravitational waves are directly observed by the Laser
Interferometer Gravitational-wave Observatory
(LIGO)~\cite{LIGO-GW150914-2016-GW151226-2016} and the
gravitational-wave astronomy has begun.
To develop this gravitational-wave astronomy as a precise science,
improvements of the detector sensitivity is necessary.
So, it is important to continue the research and development of
the science of gravitational-wave detectors together with the source
sciences of gravitational waves.
This paper is also based on such research activities.


Although some researchers already commented that the weak-value
amplification might be applicable to gravitational-wave detectors, we
have been discussed this issue, seriously.
The idea of weak measurements also proposed a new
view-point of quantum measurement theory together with an
amplification effect.
To discuss the application of this idea to gravitational-wave
detectors not only leads us to a possibility of exploring a new idea
of the gravitational-wave detection but also gives us a good
opportunity to discuss what we are doing in conventional
gravitational-wave detectors from a different view-point of quantum
measurement theory.
Therefore, it is worthwhile to discuss whether or not the idea in weak
measurements is applicable to gravitational-wave detector from many
points of view.
In particular, the comparison with conventional gravitational-wave
detectors is an important issue in such discussions.


A simple realization of the weak-value amplification is similar to the
gravitational-wave detectors in many points.
The base of the conventional gravitational-wave detectors is the Michelson
interferometer.
The arm lengths of this Michelson interferometer are tuned so that the
one of the port of the interferometer becomes the ``dark port'' as we
will explain in Sec.~\ref{sec:Michelson_weak_measurement_setup}.
Due to the propagation of gravitational waves, photons leak to the
``dark port.''
The measurement of the photon number at the ``dark port'' corresponds
to the post-selection in weak measurements.
This setup is regarded as a measurement of the effective two-level
system of the photon.
For this reason, we have been concentrated on the researches on weak
measurements for two-level
systems~\cite{K.Nakamura-A.Nishizawa-M.-K.Fujimoto-2012,A.Nishizawa-K.Nakamura-M.-K.Fujimoto-2012,K.Nakamura-M.Iinuma-2013,A.Nishizawa-2015}.
In particular, a weak-value amplification in a shot-noise limited
interferometer was
discussed~\cite{A.Nishizawa-K.Nakamura-M.-K.Fujimoto-2012}, since the
shot-noise is one of important noise in gravitational-wave detectors.


Recently, Nishizawa~\cite{A.Nishizawa-2015} reported his arguments on
the radiation-pressure noise in a weak-measurement inspired
gravitational-wave detector.
This radiation-pressure noise is also an important noise in
gravitational-wave detectors.
He also discussed ``standard quantum limit,'' which is a kind of the
sensitivity limit of the detector, and proposed an idea to break his
standard quantum limit.
Details of the detector model inspired by weak measurements in
Refs.~\cite{A.Nishizawa-K.Nakamura-M.-K.Fujimoto-2012,A.Nishizawa-2015}
will also be explained in
Sec.~\ref{sec:Michelson_weak_measurement_setup}.
In this detector model, the optical short pulse beam is used to
measure the mirror displacement due to gravitational waves, while the
continuous monochromatic laser is used for the continuous measurement
of the mirror displacement in conventional gravitational-wave
detectors.
This short-pulse injection is one of ideas in weak measurements
proposed by Aharonov et al.~\cite{Y.Aharonov-D.Z.Albert-L.Vaidman-1988}
and the main difference between a model inspired by weak measurements
in Refs.~\cite{A.Nishizawa-2015} and conventional gravitational-wave
detectors.
Furthermore, in Ref.~\cite{A.Nishizawa-2015}, arguments are
restricted to the situation where the mirror displacement is regarded
as a constant in time, while we have to monitor the motion of the
mirror displacement by the continuous laser in conventional
gravitational-wave detectors.
Due to this restriction, we cannot directly compare the results in
Ref.~\cite{A.Nishizawa-2015} with those in conventional
gravitational-wave detector and the meaning of ``standard quantum
limit'' in Ref.~\cite{A.Nishizawa-2015} is not so clear.


To monitor the time-evolution of the mirror displacement is
important in gravitational-wave detection, because it corresponds to
the monitor of the time-evolution of gravitational waves.
Expected gravitational-wave signals are in the frequency range
from 10~Hz to 10~kHz.
When we apply the detector model in Ref.~\cite{A.Nishizawa-2015}, we
may inject femto-second pulses into the interferometer and a
sufficiently large number of pulses are used to measure 10~kHz
signals.
Since we want to continuously measure the time-evolution of
gravitational-wave signal in the range 10~Hz-10~kHz,
we have to evaluate the averaged data of many pulses, continuously.
To accomplish this averaged measurement, the different treatment of
the detector in Ref.~\cite{A.Nishizawa-2015} is required.
In conventional gravitational-wave detectors, the response of the
detector to gravitational waves is discussed through the input-output
relation of the interferometer in the frequency domain in the range of
frequencies of
gravitational-waves~\cite{H.J.Kimble-Y.Levin-A.B.Matsko-K.S.Thorne-S.P.Vyatchanin-2001}.
Therefore, to compare the results with conventional gravitational-wave
detectors, it is natural to discuss the input-output relation for the
weak-measurement inspired detector model in
Ref.~\cite{A.Nishizawa-2015} taking into account of the
time-dependence of the mirror displacement.


In this paper, we regard that the Fourier transformation of optical
fields are averaged variable of many pulses with the time scale
which cover the appropriate frequency range and derive the
input-output relation for the model in Ref.~\cite{A.Nishizawa-2015}.
The important motivation of this extension is the comparison with the
conventional gravitational-wave detectors.
To carry out this extension, we have to consider at least two issues.
The first issue is in the formulation to describe the input-output
relation for the interferometers.
In the conventional gravitational-wave detectors, the input-output
relations are always derived through the two-photon formulation
developed by Caves and Schumaker~\cite{C.M.Caves-B.L.Schumaker-1985}.
However, it is not clear whether or not this two-photon formulation
can be applicable to the situation of weak measurements, because the
aim of this formulation is to discuss the sideband fluctuations at
the frequency $\omega_{0}\pm\Omega$ where $\omega_{0}$
is the frequency of the monochromatic laser and $\Omega$ is the
frequency of fluctuations around this monochromatic laser.
Furthermore, we consider the situation $\omega_{0}\gg\Omega$ in the
two-photon formulation.
It is not clear whether or not the situation $\omega_{0}\gg\Omega$ is
appropriate for the model in Ref.~\cite{A.Nishizawa-2015}.
We also note that there is little literature in which the input-output
relation for gravitational-wave detectors is derived without the
two-photon formulation.
Therefore, we have to re-derive the input-output relations of
gravitational-wave detectors without using the two-photon
formulation from the starting point.


The second issue is the extension of the photon state from the light
source in the interferometer.
In conventional gravitational-wave detectors, the optical field from
the light source is in the coherent state whose complex amplitude is
given by the $\delta$-function in the frequency domain.
On the other hand, the photon state from the light source in the model
of Ref.~\cite{A.Nishizawa-2015} is also a coherent state but its
complex amplitude has the broad band support in the frequency domain,
which corresponds to the optical pulse.
Therefore, we extend the input-output relation for conventional
gravitational-wave detectors to the situation where the state of the
injected light source is in an arbitrary coherent state.
This extension is the main purpose of this paper.
As the result of this extension, we can treat the situation of
conventional gravitational-wave detectors and that of the
model in Ref.~\cite{A.Nishizawa-2015} from the same input-output
relation and compare these models.
Furthermore, we can easily see that conventional gravitational-wave
detectors already and implicitly includes the essence of the
weak-value amplification as the noise reduction from the light source
through the measurement at the dark port.


This paper is organized as follows.
In Sec.~\ref{sec:Michelson_weak_measurement_setup}, we explain the
setup of a simple conventional Michelson gravitational-wave detector
and its weak-measurement inspired version discussed in
Ref.~\cite{A.Nishizawa-2015}.
In Sec.~\ref{sec:Extension_of_Input-output}, we derive the generalized
input-output relation which is applicable to the situation where the
injected optical beam is an arbitrary coherent state and the derived
input-output relation is the main result of this paper.
In Sec.~\ref{sec:Re-derivation_of_conventional_input-output_relation},
we re-derive the conventional input-output relation from our derived
extended input-output relation in
Sec.~\ref{sec:Extension_of_Input-output}, which indicates that our
extended input-output relation is a natural extension of the
input-output relation for conventional gravitational-wave detectors.
In Sec.~\ref{sec:WVA_from_the_extended_input-output_relation}, we
discuss the situation of the weak-value amplification of the model in
Ref.~\cite{A.Nishizawa-2015} from our derived input-output relation in
Sec.~\ref{sec:Extension_of_Input-output}, which actually realizes the
weak-value amplification.
Final section, Sec.~\ref{sec:Summary_Discussion}, is devoted to
summary and discussion which includes the comparison of the model in
Ref.~\cite{A.Nishizawa-2015} and the conventional Michelson
gravitational-wave detector.


\section{Michelson weak measurement setup}
\label{sec:Michelson_weak_measurement_setup}

\begin{figure}
  \centering
  \includegraphics[width=0.85\textwidth]{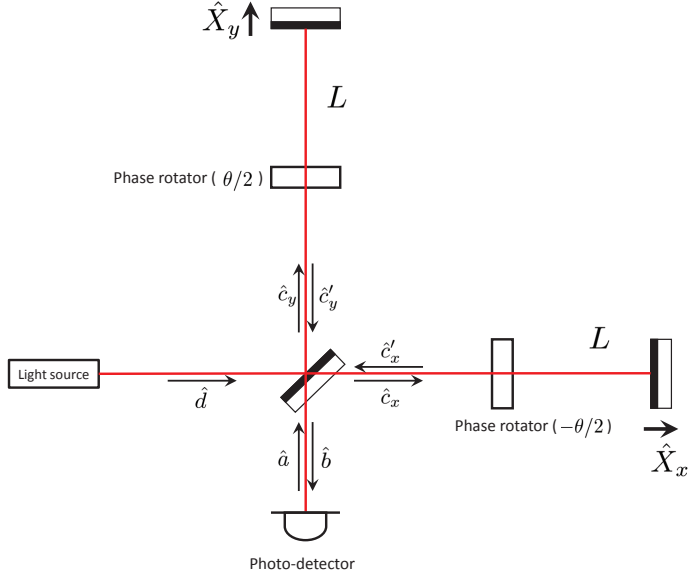}
  \caption{
    The interferometer setup for the Michelson gravitational-wave
    detector.
    The central beam splitter splits the optical beam to the $x$  and
    $y$ directions and end-mirrors reflect these optical beams to the
    beam splitter.
    The geodesic distance from the beam splitter and end-mirrors are
    almost tuned as $L$ but the mirrors can move this central
    distance $L$ with tiny distances $\hat{X}_{x}$ and  $\hat{X}_{y}$,
    respectively, along the longitudinal direction of optical beams.
    Through this setup, we want to measure the effects due to the tiny
    differential motion of mirrors $\hat{X}_{x}$ and $\hat{X}_{y}$ by
    the photo-detector in the anti-symmetric port.
    In the conventional Michelson gravitational-wave detector, we
    inject the monochromatic laser from the light source in this
    interferometer setup.
    On the other hand, in the weak-measurement inspired model, we
    introduce the phase offset $\pm\theta/2$ inspired by the original
    idea of weak
    measurements~\cite{Y.Aharonov-D.Z.Albert-L.Vaidman-1988}.
    Furthermore, the injected photon field from the light source to
    the interferometer is continuous pulse-train in the
    weak-measurement model.
    The operators $\hat{a}$, $\hat{b}$, $\hat{c}_{x,y}$,
    $\hat{c}_{x,y}'$, and $\hat{d}$ describe the quadratures for the
    electric field along the beam line.
  }
  \label{fig:kouchan-Michelson-setup-notation}
\end{figure}

In this section, we explain the simplest conventional Michelson
interferometric gravitational-wave detector and its weak-measurement
inspired version discussed in Ref.~\cite{A.Nishizawa-2015}.
The interferometer setup of these two gravitational-wave detectors is
described as in Fig.~\ref{fig:kouchan-Michelson-setup-notation}.
In the setup depicted in
Fig.~\ref{fig:kouchan-Michelson-setup-notation}, the optical beam
from the light source is injected into the interferometer which
reaches to the central beam splitter.
The central beam splitter separates the optical beam into two paths.
We denote these paths as the $x$-arm and the $y$-arm, respectively.
The separated optical beams propagate along each $x$- and $y$-arms,
reach to the end-mirrors, and are reflected to the beam splitter by
these end-mirrors, again.
At the central beam splitter, a part of the reflected beams is
returned to the port at which the light source exists.
We call this port as the ``symmetric port.''
The other part of the beam goes to the port at which the
photo-detector is prepared as depicted in
Fig.~\ref{fig:kouchan-Michelson-setup-notation}.
We call this port as the ``anti-symmetric port.''


To regard the setup in Fig.~\ref{fig:kouchan-Michelson-setup-notation}
as a gravitational-wave detector, each end-mirror, which is called
$x$-end mirror and $y$-end mirror, respectively, undergoes the free-falling
motion to the longitudinal direction of the optical beam propagation,
respectively.
In general relativity, ``free-falling motions'' are called geodesic motions.
The geodesic distance from the beam splitter and each end-mirrors are
almost tuned as $L$.
We apply a proper reference
frame~\cite{C.W.Misner-T.S.Thorne-J.A.Wheeler-1973} whose center is
the central beam splitter.
When gravitational waves propagate through this interferometer, the
geodesic distances from the beam splitter to each end-mirrors are
slightly changed due to the tidal force of gravitational waves.
In the proper reference frame, these tiny changes are represented by
$\hat{X}_{x}$ and $\hat{X}_{y}$ as depicted in
Fig.~\ref{fig:kouchan-Michelson-setup-notation}.
Through this setup of the interferometer, we measure the changes
$\hat{X}_{x}$ and $\hat{X}_{y}$ due to the gravitational-wave
propagation by the photo-detector at the anti-symmetric port.
Then, this setup is regarded as a gravitational-wave detector.
It is important to note that, if the additional noises other than
the gravitational-wave signals are included in these displacements
$\hat{X}_{x}$ and $\hat{X}_{y}$, we cannot distinguish the
gravitational-wave signals and these additional noises, because we
measure the gravitational-wave signal only through the mirror
displacement $\hat{X}_{x}$ and $\hat{X}_{y}$.
These setup are common both in the conventional Michelson
gravitational-wave detector and in the model discussed in
Ref.~\cite{A.Nishizawa-2015}.


\subsection{Conventional Michelson gravitational-wave detector}
\label{sec:Conventional_Michelson_GW_detector}


In the conventional gravitational-wave detector, $L$ is chosen so that
there is no photon leakage at the anti-symmetric port (the phase
offset $\theta=0$ in Fig.~\ref{fig:kouchan-Michelson-setup-notation})
when there is no gravitational-wave
propagation~\cite{phase-offset-introduction-comment}.
For this reason, the anti-symmetric port for the photo-detector is
usually called ``dark port.''
On the other hand, the symmetric port is called ``bright port.''
As mentioned above, the differential motion $\hat{X}_{x}-\hat{X}_{y}$
induced by the gravitational waves leads the leakage of photons to the
dark port.
This is the signal of the gravitational-wave detection.


In the conventional Michelson interferometric gravitational-wave
detector, the state of the electric field from the light source is the
monochromatic continuous laser.
In quantum theory, this state is characterized by the coherent state
with the $\delta$-function complex amplitude with the carrier
frequency $\omega_{0}$ in the frequency domain.


\subsection{Weak-measurement-inspired version}
\label{sec:Weak-measurement-ispired_version_of_Michelson_GW_detector}


Now, we explain the gravitational-wave detector in
Fig.~\ref{fig:kouchan-Michelson-setup-notation} from the view point of
the standard explanation of weak
measurements~\cite{A.Nishizawa-2015}.
The photon states propagate $x$- and $y$-arms are denoted by
$|x\rangle$ and $|y\rangle$, respectively, and the system to be
measured in quantum measurement theory is the which-path information
which is a two-level system spaned by the basis
$\{|x\rangle,|y\rangle\}$.
The difference $\hat{X}_{x}-\hat{X}_{y}$, which includes the
gravitational-wave signal, is regarded as the interaction strength
between the system and the meter variable in quantum measurement
theory.
We regard that this interaction affects the photon state at the time
$t=t_{0}$ of the reflection at the end mirrors.
The meter variable in this setup is the frequency of photon in the interferometer.


In contrast to the conventional Michelson gravitational-wave detector,
we introduce the relative phase offset $\pm\theta/2$ in each
arm~\cite{phase-offset-introduction-comment} in the weak-measurement
version of this detector.
This introduction of the phase offset is inspired by the original idea
of weak measurements~\cite{Y.Aharonov-D.Z.Albert-L.Vaidman-1988} as
explained in Ref.~\cite{A.Nishizawa-2015}.
The weak value amplification occurs when this relative phase offset is
approaching to vanishing as discussed in
Ref.~\cite{Y.Aharonov-D.Z.Albert-L.Vaidman-1988}.
Furthermore, we assume that the state of the electric field from the
light source is a coherent state whose complex amplitude in the
frequency domain has a broad band support, while the $\delta$-function
complex amplitude at the carrier frequency $\omega_{0}$ is used in the
conventional gravitational-wave detectors.
In other words, in the weak-measurement version, the pulse light is
injected into the interferometer instead of the monochromatic laser.
This is due to the fact that the meter variable to observe the
interaction strength $\hat{X}_{x}-\hat{X}_{y}$ is the frequency
distribution of photon and the variance of the meter variable should
be large in the original idea of weak
measurements~\cite{Y.Aharonov-D.Z.Albert-L.Vaidman-1988}.


Due to the above phase offset $\theta$, the initial state of the
photon is given by
\begin{eqnarray}
  \label{eq:standard-photon-preselecton-state}
  |\psi_{i}\rangle
  =
  \frac{1}{\sqrt{2}} \left(
  e^{i\theta/2} |y\rangle + e^{-i\theta/2}|x\rangle
  \right)
  .
\end{eqnarray}
This state corresponds to the photon state propagating from the
central beam splitter to the end mirrors.
On the other hand, the initial state of the photon meter variable is
given by
\begin{eqnarray}
  \label{eq:standard-total-instate}
  |\Phi\rangle = \int dp \Phi(p) |p\rangle
\end{eqnarray}
where $p$ is the momentum, or equivalently photon frequency in the
natural unit $c=1$.
The pre-selected state for the total system is $|\psi_{i}\rangle|\Phi\rangle$.
In the situation where the interaction strength
$\hat{X}_{x}-\hat{X}_{y}=: -g$ is almost constant, the reflection at
the end mirrors at the moment $t=t_{0}$ changes the state of photons
through the interaction Hamiltonian
\begin{eqnarray}
  \label{eq:standard-weak-interaction}
  \hat{H} = g\delta(t-t_{0}) \hat{A}\otimes p,
\end{eqnarray}
where and $\hat{A}$ is the which-path operator
\begin{eqnarray}
  \label{eq:standard-which-path-operator}
  \hat{A} := |y\rangle\langle y| - |x\rangle\langle x|.
\end{eqnarray}
After the interaction (\ref{eq:standard-weak-interaction}), we perform
the post-selection to the which-path information
\begin{eqnarray}
  \label{eq:standard-post-selection}
  |\psi_{f}\rangle
  =
  \frac{1}{\sqrt{2}}\left(|y\rangle-|x\rangle\right).
\end{eqnarray}
This corresponds to the detection of the photon at the anti-symmetric
port of the interferometer depicted in
Fig.~\ref{fig:kouchan-Michelson-setup-notation}.


The weak value for this measurement model is given by
\begin{eqnarray}
  \label{eq:standard-weak-value}
  A_{w}
  :=
  \frac{
  \langle\psi_{f}|\hat{A}|\psi_{i}\rangle
  }{
  \langle\psi_{f}|\psi_{i}\rangle
  }
  =
  - i \cot\frac{\theta}{2}
  .
\end{eqnarray}
The final state of the output photon after the post-selection is given
by
\begin{eqnarray}
  |\Phi'\rangle
  &:=&
  \int dp \Phi'(p)|p\rangle
  =
  \langle\psi_{f}|e^{-ig\hat{A}\otimes p}|\psi_{i}\rangle|\Phi\rangle
       \nonumber\\
  &=&
  \int dp \Phi(p)|p\rangle (1 - i A_{w} gp) + O(g^{2})
  .
  \label{eq:standard-total-outstate}
\end{eqnarray}
Evaluating the expectation value of the momentum $p$ under this final
state, we obtain
\begin{eqnarray}
  \label{eq:standard-p-expectation}
  \langle p\rangle'
  -
  \langle p\rangle
  \sim
  2 g \mbox{Im}A_{w} \left(\langle p^{2}\rangle-\langle p\rangle^{2}\right)
\end{eqnarray}
If we apply the standing point that we want to measure the interaction
strength $g:=\hat{X}_{y}-\hat{X}_{x}$ as a gravitational-wave
detector, Eq.~(\ref{eq:standard-p-expectation}) shows that the output
is proportional to $g\mbox{Im}A_{w}\propto g\cot(\theta/2)$.
If we consider the situation $\theta\ll 1$, the interaction strength
$g:=\hat{X}_{y}-\hat{X}_{x}$ can be measured by the large factor
$\sim 1/\theta$.
This is the original argument of weak-value amplification in the case
of the imaginary weak value.


The above explanation can easily be extended to multiple
photons~\cite{A.Nishizawa-2015}.
The probability distribution for a single photon at the output is
given through Eq.~(\ref{eq:standard-total-outstate}) as
\begin{eqnarray}
  \label{eq:standard-photon-frequency-distribution}
  \rho(\omega)
  :=
  \frac{
  \langle\omega|\Phi'\rangle\langle\Phi'|\omega\rangle
  }{
  \langle\Phi'|\Phi'\rangle
  }
  .
\end{eqnarray}
When the total photon number is $N_{out}$, the photon-number
distribution is simply given by
\begin{eqnarray}
  \label{eq:standard-many-photon-frequency-distribution}
  \overline{n(\omega)} = N_{out} \rho(\omega).
\end{eqnarray}
This implies that the probability distribution $\rho(\omega)$ is
regarded as the normalized photon number distribution $f(\omega)$ in
the frequency domain defined by~\cite{A.Nishizawa-2015}
\begin{eqnarray}
  \label{eq:f-distribution-def}
  f(\omega) := \frac{\overline{n(\omega)}}{N_{out}}
  :=
  \frac{
  \displaystyle
  \overline{n(\omega)}
  }{
  \displaystyle
  \int_{0}^{\infty} d\omega\bar{n}(\omega)
  }
  .
\end{eqnarray}
In the Heisenberg picture, the output photon number is given by the
expectation value of the number operator
$\hat{n}(\omega)=\hat{b}^{\dagger}(\omega)\hat{b}(\omega)$ through the
output annihilation operator $\hat{b}(\omega)$ which is introduced in
Fig.~\ref{fig:kouchan-Michelson-setup-notation}.
This will be discussed in
Sec.~\ref{sec:WVA_from_the_extended_input-output_relation} after
derived the input-output relation for the interferometer setup which
are applicable to the situation of the weak measurement.




In the understanding of conventional gravitational-wave detectors,
the input-output relation for the interferometer plays crucial
role~\cite{H.J.Kimble-Y.Levin-A.B.Matsko-K.S.Thorne-S.P.Vyatchanin-2001}.
This input-output relation is based on the quantum field theory of the
photon field.
However, the weak-measurement-inspired version is not described by the
conventional input-output relation of gravitational-wave detectors.
In this paper, we want to discuss these two typical situation
within the same mathematical framework.
To carry out this, we have to extend the conventional input-output
relation to the situation where the coherent state from the light
source has an arbitrary complex amplitude in the frequency domain as
shown in the next section.


\section{Extension of Input-output relation to arbitrary coherent state}
\label{sec:Extension_of_Input-output}


In this section, we derive an extension of the input-output relation
of the Michelson gravitational-wave detector, which is depicted in
Fig.~\ref{fig:kouchan-Michelson-setup-notation}, to an arbitrary
coherent state light source in terms of the one-photon formulation.
In many literature of gravitational-wave detectors, the two-photon
formulation developed by Caves and
Schumaker~\cite{C.M.Caves-B.L.Schumaker-1985} is used.
However, as emphasized in Sec.~\ref{sec:introduction}, it is not clear
whether or not this two-photon formulation is applicable to the
situation of the detector model in Ref.~\cite{A.Nishizawa-2015}.
Therefore, we reexamine the derivation of the input-output relation
from the starting point.
This section is the main ingredient of this paper.


In Sec.~\ref{sec:One-photon_formulation-1}, we describe the notation
of the electric field in the interferometer.
In Sec.~\ref{sec:Extension_of_Input-output_relation}, we derive the
input-output relation of the interferometer in which the photon state
from the light source is an arbitrary coherent state.
In Sec.~\ref{sec:Remarks}, we summarize remarks on the result in this
section.


\subsection{Electric field notation}
\label{sec:One-photon_formulation-1}


As well known, the electric field operator at time $t$ and the length
of the propagation direction $z$ in interferometers is described by
\begin{eqnarray}
  \label{eq:electric_field_original-1}
  \hat{E}_{a}(t-z)
  &=&
  \hat{E}^{(+)}_{a}(t-z) + \hat{E}^{(-)}_{a}(t-z)
  ,\\
  \label{eq:electric_field_original-2}
  \hat{E}_{a}^{(-)}(t-z)
  &=&
  \left[\hat{E}_{a}^{(+)}(t-z)\right]^{\dagger},
  \\
  \label{eq:electric_field_original-3}
  \hat{E}^{(+)}_{a}(t-z)
  &=&
  \int_{0}^{\infty}
      \frac{d\omega}{2\pi}
      \sqrt{\frac{2\pi\hbar\omega}{{\cal A}c}}
      \hat{a}(\omega) e^{-i\omega(t-z)}
      ,
\end{eqnarray}
where $\hat{a}(\omega)$ is the photon annihilation operator associated
with the electric field $\hat{E}_{a}(t-z)$, which satisfies the
commutation relations
\begin{eqnarray}
  \label{eq:hata-commutation-relations-1}
  \left[\hat{a}(\omega), \hat{a}^{\dagger}(\omega')\right]
  =
  2 \pi \delta(\omega-\omega')
  , \\
  \label{eq:hata-commutation-relations-}
  \left[\hat{a}(\omega), \hat{a}(\omega')\right]
  =
  \left[\hat{a}^{\dagger}(\omega), \hat{a}^{\dagger}(\omega')\right]
  =
  0
  .
\end{eqnarray}
${\cal A}$ is the cross-sectional area of the optical beam.
To discuss the input-output relation of the Michelson interferometer
based on one-photon formulation, it is convenient to introduce
operator $\hat{A}(\omega)$ defined by
\begin{eqnarray}
  \label{eq:hatA-operator-def}
  \hat{A}(\omega)
  :=
  \hat{a}(\omega) \Theta(\omega)
  +
  \hat{a}^{\dagger}(-\omega) \Theta(-\omega)
\end{eqnarray}
so that the electric field (\ref{eq:electric_field_original-1}) is
represented as
\begin{eqnarray}
  \hat{E}_{a}(t)
  =
  \int_{-\infty}^{+\infty}
  \frac{d\omega}{2\pi}
  \sqrt{\frac{2\pi\hbar|\omega|}{{\cal A}c}}
  \hat{A}(\omega)
  e^{-i\omega t}
  ,
  \label{eq:electric_field_kouchan_notation}
\end{eqnarray}
where $\Theta(\omega)$ is the Heaviside step function
\begin{eqnarray}
  \label{eq:Hevisite-step-function-def}
  \Theta(\omega)
  =
  \left\{
    \begin{array}{lcl}
      1 & \quad & (\omega\geq 0), \\
      0 & \quad & (\omega<0).
    \end{array}
  \right.
\end{eqnarray}
Due to the property of the Dirac $\delta$-function
$\int_{-\infty}^{+\infty}dt e^{+i(\omega'-\omega)t}$ $=$ $2\pi
\delta(\omega'-\omega)$, the inverse relation of
Eq.~(\ref{eq:electric_field_kouchan_notation}) is given by
\begin{eqnarray}
  \hat{A}(\omega)
  &=&
  \sqrt{\frac{{\cal A}c}{2\pi\hbar|\omega|}}
  \int_{-\infty}^{+\infty} dt e^{+i\omega t} \hat{E}_{a}(t)
  .
  \label{eq:electric_field_kouchan_notation_inverse}
\end{eqnarray}
Therefore, the operator $\hat{A}(\omega)$ includes complete
information of the electric field operator $\hat{E}_{a}(t)$ and is
convenient to derive the input-output relation of simple
interferometers.


\subsection{Input-output relation for the extended Michelson
  interferometer}
\label{sec:Extension_of_Input-output_relation}


In this subsection, we consider the extension of the input-output
relation for the Michelson interferometer depicted in
Fig.~\ref{fig:kouchan-Michelson-setup-notation}.
This ``extension'' includes three meanings.
First, the state of the input optical field is unspecified, while that
is a coherent state whose complex amplitude is a real
$\delta$-function in the frequency domain in conventional
gravitational-wave detectors.
Second, the tiny motions $\hat{X}_{x}$ and $\hat{X}_{y}$ of
end-mirrors are not specified within this subsection.
Finally, we introduce the phase offset $\pm\theta/2$ for each arm as
depicted in Fig.~\ref{fig:kouchan-Michelson-setup-notation}.
This phase offset is inspired by the original idea of weak
measurements~\cite{Y.Aharonov-D.Z.Albert-L.Vaidman-1988,phase-offset-introduction-comment}.


\subsubsection{Beam splitter junctions}
\label{sec:Beam_splitter_junction}


First, we consider the junction conditions for quadratures at the beam
splitter.
Following the notation depicted in
Fig.~\ref{fig:kouchan-Michelson-setup-notation}, the final output
electric field $\hat{E}_{b}(t)$ is given
by~\cite{beam-splitter-comment}
\begin{eqnarray}
  \label{eq:output-electric-field-junction}
  \hat{E}_{b}(t)
  =
  \frac{1}{\sqrt{2}} \left[
    \hat{E}_{c_{y}'}(t) - \hat{E}_{c_{x}'}(t)
  \right]
  .
\end{eqnarray}
Here, we define
\begin{eqnarray}
  \hat{B}(\omega)
  &:=&
  \hat{b}(\omega) \Theta(\omega)
  +
  \hat{b}^{\dagger}(-\omega) \Theta(-\omega)
  \label{eq:hatB-def}
  , \\
  \hat{C}_{x}'(\omega)
  &:=&
  \hat{c}_{x}'(\omega) \Theta(\omega)
  +
  \hat{c}_{x}^{'\dagger}(-\omega) \Theta(-\omega)
  \label{eq:hatCx'-def}
  , \\
  \hat{C}_{y}'(\omega)
  &:=&
  \hat{c}_{y}'(\omega) \Theta(\omega)
  +
  \hat{c}_{y}^{'\dagger}(-\omega) \Theta(-\omega)
  \label{eq:hatCy'-def}
\end{eqnarray}
as in Eq.~(\ref{eq:hatA-operator-def}) and the relation
(\ref{eq:output-electric-field-junction}) is given by
\begin{eqnarray}
  \label{eq:output-electric-field-junction-hatB}
  \hat{B}(\omega)
  =
  \frac{1}{\sqrt{2}} \left(
    \hat{C}_{y}'(\omega) - \hat{C}_{x}'(\omega)
  \right),
\end{eqnarray}
Similarly, the electric-field operators $\hat{E}_{c_{y}}(t)$ and
$\hat{E}_{c_{x}}(t)$ are also given by the input fields
$\hat{E}_{d}(t)$ and $\hat{E}_{a}(t)$ as follows~\cite{beam-splitter-comment}:
\begin{eqnarray}
  \label{eq:xarm-input-field-junction}
  \hat{E}_{c_{x}}(t)
  &=&
  \frac{1}{\sqrt{2}} \left(
    \hat{E}_{d}(t) - \hat{E}_{a}(t)
  \right)
  , \\
  \label{eq:yarm-input-field-junction}
  \hat{E}_{c_{y}}(t)
  &=&
  \frac{1}{\sqrt{2}} \left(
    \hat{E}_{d}(t) + \hat{E}_{a}(t)
  \right)
  .
\end{eqnarray}
In terms of the quadrature as in Eq.~(\ref{eq:hatA-operator-def}),
these relations yield
\begin{eqnarray}
  \label{eq:xarm-input-field-junction-hatCx}
  \hat{C}_{x}(\omega)
  &=&
  \frac{1}{\sqrt{2}} \left(
    \hat{D}(\omega) - \hat{A}(\omega)
  \right)
  , \\
  \label{eq:yarm-input-field-junction-hatCy}
  \hat{C}_{y}(\omega)
  &=&
  \frac{1}{\sqrt{2}} \left(
    \hat{D}(\omega) + \hat{A}(\omega)
  \right)
  ,
\end{eqnarray}
where we defined the operators
\begin{eqnarray}
  \hat{C}_{x}(\omega)
  &:=&
  \hat{c}_{x}(\omega) \Theta(\omega)
  +
  \hat{c}_{x}^{\dagger}(-\omega) \Theta(-\omega)
  \label{eq:hatCx-def}
  , \\
  \hat{C}_{y}(\omega)
  &:=&
  \hat{c}_{y}(\omega) \Theta(\omega)
  +
  \hat{c}_{y}^{\dagger}(-\omega) \Theta(-\omega)
  \label{eq:hatCy-def}
  , \\
  \hat{D}(\omega)
  &:=&
  \hat{d}(\omega) \Theta(\omega)
  +
  \hat{d}^{\dagger}(-\omega) \Theta(-\omega)
  \label{eq:hatD-def}
\end{eqnarray}
as in Eq.~(\ref{eq:hatA-operator-def}).


\subsubsection{Arm propagation}
\label{sec:Arm_propagations}


Next, we consider the retarded effect due to the propagation along
each each $x$- and $y$-arm.
Here, each arm length is given by $L=c\tau$, where $\tau$ is the
retarded time for photons which propagate from the beam splitter to
the end-mirrors.
In addition to the retarded time $\tau$, the tiny displacement of the
mirror $\hat{X}_{x}/c$ and $\hat{X}_{y}/c$ also contribute to the
phase shift of the electric field.
In addition to these retarded effects, we add the retarted time
$\Delta t_{\theta}$ which corresponds to the phase offset
$\pm\theta/2$ in Fig.~\ref{fig:kouchan-Michelson-setup-notation}.
Then, the relations between the electric field
$\{\hat{E}_{c_{x}'}(t),\hat{E}_{c_{y}'}(t)\}$ and
$\{\hat{E}_{c_{x}}(t),\hat{E}_{c_{y}}(t)\}$ are given by
\begin{eqnarray}
  \label{eq:xarm-retarded-effect}
  \hat{E}_{c'_{x}}(t)
  &=&
  \hat{E}_{c_{x}}[t - 2(\tau + \hat{X}_{x}(t)/c)+\Delta t_{\theta}]
  , \\ 
  \label{eq:yarm-retarded-effect}
  \hat{E}_{c'_{y}}(t)
  &=&
  \hat{E}_{c_{y}}[t - 2(\tau + \hat{X}_{y}(t)/c)-\Delta t_{\theta}].
\end{eqnarray}
where
\begin{eqnarray}
  \label{eq:Delta-t-theta-def}
  \frac{\theta}{2} = \omega\Delta t_{\theta}(\omega).
\end{eqnarray}


Here, we treat $\hat{X}_{x}/c$ and $\hat{X}_{y}/c$, perturbatively.
Through the representation of the electric fields $\hat{E}_{c_{x}}(t)$
and $\hat{E}_{c'_{x}}(t)$ as
Eq.~(\ref{eq:electric_field_kouchan_notation}), the relation
(\ref{eq:xarm-retarded-effect}) with Eq.~(\ref{eq:Delta-t-theta-def})
is given by
\begin{eqnarray}
  \hat{E}_{c'_{x}}(t)
  &=&
  \int_{-\infty}^{+\infty}
  \frac{d\omega}{2\pi}
  \sqrt{\frac{2\pi\hbar|\omega|}{{\cal A}c}}
  \hat{C}_{x}'(\omega)
  e^{-i\omega t}
  \nonumber\\
  &\sim&
  e^{-i\theta/2}
  \int_{-\infty}^{+\infty}
  \frac{d\omega}{2\pi}
  \sqrt{\frac{2\pi\hbar|\omega|}{{\cal A}c}}
  \hat{C}_{x}(\omega)
  e^{- i \omega (t-2\tau)}
  \nonumber\\
  &&
  +
  e^{-i\theta/2}
  \int_{-\infty}^{+\infty}
  \frac{d\omega}{2\pi}
  \sqrt{\frac{2\pi\hbar|\omega|}{{\cal A}c}}
  \hat{C}_{x}(\omega)
  e^{- i \omega (t-2\tau)}
  \frac{
    2 i \omega \hat{X}_{x}(t-\tau)
  }{
    c
  }
  .
  \label{eq:kouchan-2015-13}
\end{eqnarray}
In this expression, $\hat{X}_{x}$ is regarded as a quantum operator.
In this case, there is a ordering problem of the operators
$\hat{X}_{x}$ and $\hat{c}_{x}(\omega)$, but, here, we keep the order
in which the operator $\hat{X}_{x}$ should be in the right of the
operator  $\hat{c}_{x}(\omega)$ at this moment.
We note that this ordering problem is harmless within this paper,
because we concentrate only on the linear quadrature relations
with a coherent state light source.


Here, we introduce the Fourier transformation
\begin{eqnarray}
  \label{eq:hatXx-Fouriler-def}
  \hat{X}_{x}(t)
  &=:&
       \int_{-\infty}^{+\infty}
       \hat{Z}_{x}(\Omega)
       e^{-i\Omega t} \frac{d\Omega}{2\pi}
       ,
\end{eqnarray}
substitute Eq.~(\ref{eq:hatXx-Fouriler-def}) into
Eq.~(\ref{eq:kouchan-2015-13}), and take the Fourier transformation
(\ref{eq:electric_field_kouchan_notation_inverse}).
Then, we have
\begin{eqnarray}
  &&
  \hat{C}_{x}'(\omega)
  =
      e^{-i\theta/2}
      e^{+ 2 i \omega\tau}
      \hat{C}_{x}(\omega)
      \nonumber\\
  && \quad\quad\quad\quad
     +
     e^{-i\theta/2}
     e^{+i2\omega\tau}
     \frac{2i}{c\sqrt{|\omega|}}
     \int_{-\infty}^{+\infty}
     \frac{d\Omega}{2\pi}
     e^{-i\Omega\tau}
     \sqrt{|\omega-\Omega|}
     (\omega-\Omega)
      \nonumber\\
  && \quad\quad\quad\quad\quad\quad\quad\quad\quad\quad\quad\quad\quad\quad\quad\quad
     \times
     \hat{C}_{x}(\omega-\Omega)
     \hat{Z}_{x}(\Omega)
     .
     \nonumber\\
  \label{eq:hatC'x-hatCx-relatin}
\end{eqnarray}
Similarly, from Eq.~(\ref{eq:yarm-retarded-effect}), we
obtain
\begin{eqnarray}
  &&
  \hat{C}_{y}'(\omega)
  =
  e^{+i\theta/2}
  e^{+ 2 i \omega\tau}
  \hat{C}_{y}(\omega)
      \nonumber\\
  && \quad\quad\quad\quad
  +
  e^{+i\theta/2}
  e^{+i2\omega\tau}
  \frac{2i}{c\sqrt{|\omega|}}
  \int_{-\infty}^{+\infty}
  \frac{d\Omega}{2\pi}
  e^{-i\Omega\tau}
  \sqrt{|\omega-\Omega|}
  (\omega-\Omega)
      \nonumber\\
  && \quad\quad\quad\quad\quad\quad\quad\quad\quad\quad\quad\quad\quad\quad\quad\quad
     \times
     \hat{C}_{y}(\omega-\Omega)
     \hat{Z}_{y}(\Omega)
  \label{eq:hatC'y-hatCy-relatin}
\end{eqnarray}
through the replacements
$\hat{C}'_{x}\rightarrow\hat{C}'_{y}$,
$\hat{Z}_{x}\rightarrow\hat{Z}_{y}$ and $\theta\rightarrow-\theta$
in Eq.~(\ref{eq:hatXx-Fouriler-def}) and
(\ref{eq:hatC'x-hatCx-relatin}).


Substituting Eqs.~(\ref{eq:hatC'x-hatCx-relatin}) and
(\ref{eq:hatC'y-hatCy-relatin}) into
Eq.~(\ref{eq:output-electric-field-junction-hatB}), we obtain
\begin{eqnarray}
  &&
     \!\!\!\!\!\!\!\!\!\!\!\!\!\!\!\!\!\!
     e^{- 2 i \omega\tau} \hat{B}(\omega)
     \nonumber\\
  &=&
     i \sin\left(\frac{\theta}{2}\right)
     \hat{D}(\omega)
     +
     \cos\left(\frac{\theta}{2}\right)
     \hat{A}(\omega)
     \nonumber\\
  &&
     +
     \frac{2i}{c}
       \int_{-\infty}^{+\infty}
       \frac{d\nu}{2\pi}
       e^{-i\nu\tau}
       \sqrt{
       \left|
       \frac{\omega-\nu}{\omega}
       \right|
       }
       (\omega-\nu)
     \nonumber\\
  && \quad\quad\quad
     \times
       \left[
       \left(
       i \sin\left(\frac{\theta}{2}\right)
       \hat{D}(\omega-\nu)
       +
       \cos\left(\frac{\theta}{2}\right)
       \hat{A}(\omega-\nu)
       \right)
       \hat{Z}_{com}(\nu)
       \right.
       \nonumber\\
    && \quad\quad\quad\quad
       \left.
       -
       \left(
       \cos\left(\frac{\theta}{2}\right)
       \hat{D}(\omega-\nu)
       +
       i \sin\left(\frac{\theta}{2}\right)
       \hat{A}(\omega-\nu)
       \right)
       \hat{Z}_{diff}(\nu)
       \right],
       \nonumber\\
       \label{eq:hatB-hatD-hatA-hatZ-relation}
\end{eqnarray}
where we defined
\begin{eqnarray}
  \hat{Z}_{com}
  :=
  \frac{\hat{Z}_{x} + \hat{Z}_{y}}{2}
  , \quad
  \hat{Z}_{diff}
  :=
  \frac{\hat{Z}_{x} - \hat{Z}_{y}}{2}
  .
  \label{eq:Zdif-Zcom-def}
\end{eqnarray}


\subsubsection{Coherent state of the input optical beam}
\label{sec:Coherent_state_input_light_source}


Eq.~(\ref{eq:hatB-hatD-hatA-hatZ-relation}) indicates that the
output operator $\hat{B}$ is given by the operators $\hat{A}$,
$\hat{D}$, $\hat{Z}_{diff}$, and $\hat{Z}_{com}$.
In this section, we will see that the $\hat{Z}_{diff}$ and
$\hat{Z}_{com}$ are given by $\hat{A}$ and $\hat{D}$ together with the
gravitational-wave signal through the equations of motion for
end-mirrors, later.
Therefore, to discuss the information from the output operator
$\hat{B}$, we have to specify the quantum states associated with the
operators $\hat{A}$ and $\hat{D}$.


The state for the operator $\hat{A}$ is the state which is injected
from the anti-symmetric port.
On the other hand, the state associated with the operator $\hat{D}$ is
the state of the electric field which is injected from the symmetric
port.
The total state of photon in the output port $\hat{B}$, i.e.,
$\hat{b}$, is determined by the specification of the states for the
operators $\hat{D}$ and $\hat{A}$, i.e,  the annihilation
operators $\hat{d}$ and $\hat{a}$.
We assume that the state associated with the operator
$\hat{d}$ is a coherent state with the complex amplitude
$\alpha(\omega)$, the state associated with the operator $\hat{a}$
is vacuum state, and no entangled in these states.
Then, the total state $|\psi\rangle$ of photon is given by the direct
product of the photon states of each frequency as
\begin{eqnarray}
  |\psi\rangle
  &=&
  \prod_{\omega} |\alpha(\omega)\rangle_{d}\otimes|0\rangle_{a}
  =
  \prod_{\omega} D_{d}(\alpha(\omega))|0\rangle_{d}\otimes|0\rangle_{a}
  \nonumber\\
  &=:&
  D_{d}|0\rangle_{d}\otimes|0\rangle_{a}
  ,
  \label{eq:total-state-def}
       \\
  D_{d}
  &:=&
       \prod_{\omega} D_{d}(\alpha(\omega))
       \nonumber\\
  &=&
      \exp\left[
      \int \frac{d\omega}{2\pi}\left(
      \alpha(\omega) d^{\dagger}(\omega)
      -
      \alpha^{*}(\omega) d(\omega)
      \right)
      \right]
      .
  \label{eq:displacement-operator-for-hatd}
\end{eqnarray}


In the Heisenberg picture, the operator $\hat{d}$ is replaced as
\begin{eqnarray}
  \label{eq:coherent-state-hatd-Heisenberg}
  D_{d}^{\dagger}\hat{d}(\omega)D_{d}
  =
  \hat{d}(\omega) + \alpha(\omega)
  ,
  \\
  \label{eq:coherent-state-hatddagger-Heisenberg}
  D_{d}^{\dagger}\hat{d}^{\dagger}(\omega)D_{d}
  =
  \hat{d}^{\dagger}(\omega) + \alpha^{*}(\omega)
  .
\end{eqnarray}
In terms of the operator $\hat{D}(\omega)$ defined by
Eq.~(\ref{eq:hatD-def}), this replacement is equivalent to
\begin{eqnarray}
  \label{eq:coherent-state-hatD-Heisenberg}
  D_{d}^{\dagger}\hat{D}(\omega)D_{d}
  =
  \hat{D}_{c}(\omega) + \hat{D}_{v}(\omega),
\end{eqnarray}
where
\begin{eqnarray}
  \label{eq:hatDc-def}
  \hat{D}_{c}(\omega)
  :=
  \alpha(\omega)\Theta(\omega)
  +
  \alpha^{*}(-\omega)\Theta(-\omega)
  ,
  \\
  \label{eq:hatDv-def}
  \hat{D}_{v}(\omega)
  :=
  \hat{d}(\omega)\Theta(\omega)
  +
  \hat{d}^{\dagger}(-\omega)\Theta(-\omega)
  .
\end{eqnarray}


Since we apply the Heisenberg picture, the operator
$D_{d}^{\dagger}\hat{B}(\omega)D_{d}$ is useful for evaluation of the
photon-number expectation value from the input-output relation
(\ref{eq:hatB-hatD-hatA-hatZ-relation}).
We regard the terms
$D_{d}^{\dagger}\hat{Z}_{com}(\Omega)D_{d}$ and
$D_{d}^{\dagger}\hat{Z}_{diff}(\Omega)D_{d}$ are small correction and
we neglect the quadratic terms of these small correction.
Operating $D_{d}^{\dagger}$ and $D_{d}$ to
Eq.~(\ref{eq:hatB-hatD-hatA-hatZ-relation}), substituting
Eqs.~(\ref{eq:coherent-state-hatD-Heisenberg}) into
Eq.~(\ref{eq:hatB-hatD-hatA-hatZ-relation}), we obtain the
input-output relation as
\begin{eqnarray}
  &&
     \!\!\!\!\!\!\!\!\!\!\!\!\!\!\!\!\!\!
  e^{- 2 i \omega\tau} D_{d}^{\dagger}\hat{B}(\omega)D_{d}
     \nonumber\\
  &=&
      i \sin\left(\frac{\theta}{2}\right) \hat{D}_{c}(\omega)
  +
  i \sin\left(\frac{\theta}{2}\right) \hat{D}_{v}(\omega)
  +
  \cos\left(\frac{\theta}{2}\right) \hat{A}(\omega)
  \nonumber\\
  &&
  +
  \frac{2i}{c}
  \int_{-\infty}^{+\infty}
  \frac{d\Omega}{2\pi}
  e^{-i\Omega\tau}
  \sqrt{\left|\frac{\omega-\Omega}{\omega}\right|}
  (\omega-\Omega)
  \nonumber\\
  && \quad\quad\quad\quad
  \times
  \left[
    i \sin\left(\frac{\theta}{2}\right)
    \hat{D}_{c}(\omega-\Omega)
    D_{d}^{\dagger}\hat{Z}_{com}(\Omega)D_{d}
  \right.
  \nonumber\\
  && \quad\quad\quad\quad\quad\quad\quad
  \left.
    -
    \cos\left(\frac{\theta}{2}\right)
    \hat{D}_{c}(\omega-\Omega)
    D_{d}^{\dagger}\hat{Z}_{diff}(\Omega)D_{d}
  \right]
  .
  \label{eq:linear-hatB-coherent-state-Heisenberg}
\end{eqnarray}
This is the most general input-output relation within our
consideration.
In Eq.~(\ref{eq:linear-hatB-coherent-state-Heisenberg}), the first
term in the first line is the leakage of the classical carrier field
due to the phase offset $\theta/2$.
The remaining terms in the first line is the vacuum fluctuations which
corresponds to the shot noise.
The second- and third-lines are the response of the mirror motion which
includes gravitational-wave signal and radiation pressure noise
through the motions of mirrors
$D_{d}^{\dagger}\hat{Z}_{com}(\Omega)D_{d}$ and
$D_{d}^{\dagger}\hat{Z}_{diff}(\Omega)D_{d}$.
The input-output relation
(\ref{eq:linear-hatB-coherent-state-Heisenberg}) is the main results
of this paper.
To evaluate the input-output relation
(\ref{eq:linear-hatB-coherent-state-Heisenberg}), we have to evaluate
$D_{d}^{\dagger}\hat{Z}_{com}(\Omega)D_{d}$ and
$D_{d}^{\dagger}\hat{Z}_{diff}(\Omega)D_{d}$ in some way.


\subsubsection{End-mirrors' equations of motion (time domain)}
\label{sec:End-mirror_EOM_-time-domain}


In the case of gravitational-wave detectors,
$D_{d}^{\dagger}\hat{Z}_{com}(\Omega)D_{d}$ and
$D_{d}^{\dagger}\hat{Z}_{diff}(\Omega)D_{d}$ are evaluated through the
equations of motions for the end-mirrors.
We assume that the mass of the beam splitter and end-mirrors are equal
to $m$.
Since we apply the proper reference
frame~\cite{C.W.Misner-T.S.Thorne-J.A.Wheeler-1973} of a local
inertia system in which the beam splitter is the center of this
coordinate system, $\hat{X}_{x}$ and $\hat{X}_{y}$ describe the tiny
differential displacement of the geodesic distances of the $x$- and
$y$-end mirrors from the central beam splitter, respectively.
The equations for $\hat{X}_{x}$ and $\hat{X}_{y}$ are given by
\begin{eqnarray}
  \label{eq:eq-of-motion-xend-original}
  \frac{m}{2} \frac{d^{2}}{dt^{2}}\hat{X_{x}}(t)
  &=&
  \hat{F}_{rp(x)}(t)
  +
  \frac{1}{2} \frac{m}{2} L \frac{d^{2}}{dt^{2}}h(t)
  , \\
  \label{eq:eq-of-motion-yend-original}
  \frac{m}{2} \frac{d^{2}}{dt^{2}}\hat{X_{y}}(t)
  &=&
  \hat{F}_{rp(y)}(t)
  -
  \frac{1}{2} \frac{m}{2} L \frac{d^{2}}{dt^{2}}h(t)
  .
\end{eqnarray}
where $h$ is the gravitational wave signal which is derived from the
tidal force due to the gravitational-wave propagation in the proper
reference frame and $m/2$ is the reduced mass of the differential
motion of the end-mirrors and the central beam splitter.
Furthermore, $F_{rp(x)}$ and $F_{rp(y)}$ are the radiation pressure
due to the incident photon.


\subsubsection{Radiation pressure forces}
\label{sec:Radiation_pressure_forces}


The radiation pressure forces in
Eqs.~(\ref{eq:eq-of-motion-xend-original}) and
(\ref{eq:eq-of-motion-yend-original}) are evaluated through
\begin{eqnarray}
  \hat{F}_{rp(x)}(t)
  &=&
      2 \frac{{\cal A}}{4\pi}
      \left(
      \hat{E}_{c_{x}}
      \left[
      t
      -
      \left(\tau+\frac{\hat{X}_{x}}{c}\right)
      +
      \frac{\Delta t_{\theta}}{2}
      \right]
      \right)^{2}
      ,
      \label{eq:radiation_pressure_force_xmirror-original}
  \\
  \hat{F}_{rp(y)}(t)
  &=&
      2 \frac{{\cal A}}{4\pi}
      \left(
      \hat{E}_{c_{y}}\left[
      t
      -
      \left(\tau+\frac{\hat{X}_{y}}{c}\right)
      -
      \frac{\Delta t_{\theta}}{2}
      \right]
      \right)^{2}
      ,
      \label{eq:radiation_pressure_force_ymirror-original}
\end{eqnarray}
in this paper.
The right-hand sides in
Eqs.~(\ref{eq:radiation_pressure_force_xmirror-original}) and
(\ref{eq:radiation_pressure_force_ymirror-original}) are just twice of
the pointing flux of the electric fields which incident to the
end-mirrors, respectively.


Performing the Fourier transformation of the electric field
$\hat{E}_{c_{x}}$, using Eq.~(\ref{eq:Delta-t-theta-def}), and taking
the zeroth- and the linear-order with respect to the operator
$\hat{X}_{x}$, the radiation-pressure force
(\ref{eq:radiation_pressure_force_xmirror-original}) is given by
\begin{eqnarray}
  \hat{F}_{rp(x)}(t)
  &=&
      \frac{\hbar}{c}
      e^{-i\theta/2}
      \int_{-\infty}^{\infty}
      \int_{-\infty}^{\infty}
      \frac{d\omega}{2\pi}
      \frac{d\omega'}{2\pi}
      \sqrt{|\omega\omega'|}
      \hat{C}_{x}(\omega)
      \hat{C}_{x}(\omega')
      e^{+i(\omega+\omega')\tau}
      e^{-i(\omega+\omega')t}
      \nonumber\\
  &&
     +
     i
     \frac{\hbar}{c}
     e^{-i\theta/2}
     \int_{-\infty}^{\infty}
     \int_{-\infty}^{\infty}
     \frac{d\omega}{2\pi}
     \frac{d\omega'}{2\pi}
     \sqrt{|\omega\omega'|}
     (\omega+\omega')
     \hat{C}_{x}(\omega)
     \hat{C}_{x}(\omega')
     \nonumber\\
  && \quad\quad\quad\quad\quad\quad\quad\quad\quad\quad\quad
     \times
     \frac{\hat{X}_{x}(t)}{c}
     e^{+i(\omega+\omega')\tau}
     e^{-i(\omega+\omega')t}
     .
     \label{eq:hatFrpx-time-domain-original}
\end{eqnarray}
Substituting the Fourier transformation (\ref{eq:hatXx-Fouriler-def})
of  $\hat{X}_{x}$ and taking Fourier transformation
$\hat{F}_{rp(x)}(\Omega)$ of $\hat{F}_{rp(x)}(t)$, we obtain the
expression of the radiation-pressure force which affect to the $x$-end
mirror in the frequency domain as
\begin{eqnarray}
  \hat{F}_{rp(x)}(\Omega)
  &:=&
       \int_{-\infty}^{+\infty} dt
       \hat{F}_{rp(x)}(t)
       e^{+i\Omega t}
       \nonumber\\
  &=&
      \frac{\hbar}{c}
      e^{-i\theta/2}
      e^{+i\Omega\tau}
      \int
      \frac{d\omega}{2\pi}
      \sqrt{|\omega(\Omega-\omega)|}
      \hat{C}_{x}(\omega)
      \hat{C}_{x}(\Omega-\omega)
      \nonumber\\
  &&
     +
     i
     \frac{\hbar}{c^{2}}
     e^{-i\theta/2}
     \iint
     \frac{d\omega}{2\pi}
     \frac{d\omega'}{2\pi}
     \sqrt{|\omega\omega'|}
     (\omega+\omega')
     \hat{C}_{x}(\omega)
     \hat{C}_{x}(\omega')
     \nonumber\\
  && \quad\quad\quad\quad\quad\quad\quad\quad\quad
     \times
     \hat{Z}_{x}(\Omega-\omega-\omega')
     e^{+i(\omega+\omega')\tau}
     ,
     \label{eq:hatFrpx-Fourier-domain-original}
\end{eqnarray}
where we use the notation $\int=\int_{-\infty}^{+\infty}$ and
$\iint=\int_{-\infty}^{+\infty}\int_{-\infty}^{+\infty}$.
Similarly, we also obtain
\begin{eqnarray}
  \hat{F}_{rp(y)}(\Omega)
  &:=&
       \int_{-\infty}^{+\infty} dt
       \hat{F}_{rp(y)}(t)
       e^{+i\Omega t}
       \nonumber\\
  &=&
      \frac{\hbar}{c}
      e^{+i\theta/2}
      e^{+i\Omega\tau}
      \int
      \frac{d\omega}{2\pi}
      \sqrt{|\omega(\Omega-\omega)|}
      \hat{C}_{y}(\omega)
      \hat{C}_{y}(\Omega-\omega)
      \nonumber\\
  &&
     +
     i
     \frac{\hbar}{c^{2}}
     e^{+i\theta/2}
     \iint
     \frac{d\omega}{2\pi}
     \frac{d\omega'}{2\pi}
     \sqrt{|\omega\omega'|}
     (\omega+\omega')
     \hat{C}_{y}(\omega)
     \hat{C}_{y}(\omega')
     \nonumber\\
  && \quad\quad\quad\quad\quad\quad\quad\quad\quad
     \times
     \hat{Z}_{y}(\Omega-\omega-\omega')
     e^{+i(\omega+\omega')\tau}
     .
     \label{eq:hatFrpy-Fourier-domain-original}
\end{eqnarray}


\subsubsection{End-mirrors' equations of motions (Frequency domain)} 
\label{sec:End-mirror_EOM_-Frequency-domain}


$D_{d}^{\dagger}\hat{Z}_{com}(\Omega)D_{d}$ and
$D_{d}^{\dagger}\hat{Z}_{diff}(\Omega)D_{d}$ in the input-output
relation (\ref{eq:linear-hatB-coherent-state-Heisenberg}) are
determined by Eqs.~(\ref{eq:eq-of-motion-xend-original}) and
(\ref{eq:eq-of-motion-yend-original}) of motions for the test masses
in the frequency domain.
Multiplying $D_{d}^{\dagger}$ and $D_{d}$ to the Fourier transformed
version of Eqs.~(\ref{eq:eq-of-motion-xend-original}) and
(\ref{eq:eq-of-motion-yend-original}), we obtain
\begin{eqnarray}
  m \Omega^{2}
  D_{d}^{\dagger}
  \hat{Z}_{com}(\Omega)
  D_{d}
  &=&
  -
  D_{d}^{\dagger}
  \hat{F}_{rp(x)}(\Omega)
  D_{d}
  -
  D_{d}^{\dagger}
  \hat{F}_{rp(y)}(\Omega)
  D_{d}
  ,
  \label{eq:DDdaggered-eq-of-motion-com-freq.-domain}
      \\
  m \Omega^{2}
  D_{d}^{\dagger}
  \hat{Z}_{diff}(\Omega)
  D_{d}
  &=&
  D_{d}^{\dagger}
  \hat{F}_{rp(y)}(\Omega)
  D_{d}
  -
  D_{d}^{\dagger}
  \hat{F}_{rp(x)}(\Omega)
  D_{d}
  \nonumber\\
  &&
  +
  \frac{1}{2}
  m L
  \Omega^{2}
  h(\Omega)
  \label{eq:DDdaggered-eq-of-motion-dif-freq.-domain}
\end{eqnarray}
from definitions (\ref{eq:Zdif-Zcom-def}) of $\hat{Z}_{com}(\Omega)$
and $\hat{Z}_{diff}(\Omega)$.
Here, we have regarded that the gravitational-wave signal
$h(\Omega)$ is a classical variable which is proportional to
the identity operator in the sense of quantum theory.
We also used the displacement operator $D_{d}$ is time-independent.
Equations
(\ref{eq:DDdaggered-eq-of-motion-com-freq.-domain}) and
(\ref{eq:DDdaggered-eq-of-motion-dif-freq.-domain}) indicate that we
have to evaluate $D_{d}^{\dagger}\hat{F}_{rp(x)}(\Omega)D_{d}$ and
$D_{d}^{\dagger}\hat{F}_{rp(y)}(\Omega)D_{d}$ to evaluate
$D_{d}^{\dagger}\hat{Z}_{com}(\Omega)D_{d}$ and
$D_{d}^{\dagger}\hat{Z}_{diff}(\Omega)D_{d}$.


Note that $D_{d}^{\dagger}\hat{C}_{x}(\omega)D_{d}$ in
$D_{d}^{\dagger}\hat{F}_{rp(x)}(\Omega)D_{d}$ is given by the
quadrature $\hat{D}(\omega)$ and $\hat{A}(\omega)$ through
Eq.~(\ref{eq:xarm-input-field-junction-hatCx}).
We consider the situation where the state for the input
quadrature $\hat{A}(\omega)$ from the anti-symmetric port is vacuum
and the state for the input quadrature $\hat{D}(\omega)$ from the
symmetric port is the coherent state as
Eq.~(\ref{eq:total-state-def}) which enable us to separate the
operator $\hat{D}(\omega)$ into the vacuum quadrature and the
classical carrier as Eq.~(\ref{eq:coherent-state-hatD-Heisenberg}).
Through Eqs.~(\ref{eq:xarm-input-field-junction-hatCx}),
(\ref{eq:yarm-input-field-junction-hatCy}) and
(\ref{eq:coherent-state-hatD-Heisenberg}), we may separate
$D_{d}^{\dagger}\hat{C}_{x,y}(\omega)D_{d}$ into the vacuum
quadrature and the classical carrier as
\begin{eqnarray}
  \label{eq:DDdaggered-hatCxy-separation}
  D_{d}^{\dagger}\hat{C}_{x,y}(\omega)D_{d}
  =
  \hat{C}_{x,y(c)}(\omega) + \hat{C}_{x,y(v)}(\omega),
\end{eqnarray}
where
\begin{eqnarray}
  \label{eq:hatCx,y(c)-def}
  \hat{C}_{x(c)}(\omega)
  &=&
  \hat{C}_{y(c)}(\omega)
  =
  \frac{1}{\sqrt{2}} \hat{D}_{c}(\omega)
      ,
      \\
  \label{eq:hatCx(v)-def}
  \hat{C}_{x(v)}(\omega)
  &:=&
  \frac{1}{\sqrt{2}} \left(
    \hat{D}_{v}(\omega)
    + \hat{A}(\omega)
  \right)
  ,
      \\
  \label{eq:hatCy(v)-def}
  \hat{C}_{y(v)}(\omega)
  &:=&
  \frac{1}{\sqrt{2}} \left(
    \hat{D}_{v}(\omega)
    -
    \hat{A}(\omega)
  \right)
  .
\end{eqnarray}


Through Eqs.~(\ref{eq:hatFrpx-Fourier-domain-original}),
(\ref{eq:hatFrpy-Fourier-domain-original}),
(\ref{eq:DDdaggered-hatCxy-separation}), (\ref{eq:hatCx,y(c)-def}),
(\ref{eq:hatCx(v)-def}), and (\ref{eq:hatCy(v)-def}), and ignoring the
higher-order terms of the vacuum quadrature and displacement, we
obtain
\begin{eqnarray}
  &&
     \!\!\!\!\!\!\!\!\!\!\!\!\!\!\!\!\!\!
  D_{d}^{\dagger}\hat{F}_{rp(x)}(\Omega)D_{d}
     \nonumber\\
  &=&
      \frac{\hbar}{2c}
      e^{-i\theta/2}
      e^{+i\Omega\tau}
      \int
      \frac{d\omega}{2\pi}
      \sqrt{|\omega(\Omega-\omega)|}
      \left[
      \hat{D}_{c}(\omega)
      \hat{D}_{c}(\Omega-\omega)
      \right.
      \nonumber\\
  && \quad\quad\quad\quad\quad\quad\quad\quad\quad
     \left.
     +
     2
     \hat{D}_{c}(\omega)
     \left(
     \hat{D}_{v}(\Omega-\omega)
     + \hat{A}(\Omega-\omega)
     \right)
     \right]
     \nonumber\\
  &&
     +
     i
     \frac{\hbar}{2c^{2}}
     e^{-i\theta/2}
     \iint
     \frac{d\omega}{2\pi}
     \frac{d\omega'}{2\pi}
     \sqrt{|\omega\omega'|}
     (\omega+\omega')
     \hat{D}_{c}(\omega)
     \hat{D}_{c}(\omega')
     \nonumber\\
  && \quad\quad\quad\quad\quad\quad\quad\quad\quad
     \times
     D_{d}^{\dagger}\hat{Z}_{x}(\Omega-\omega-\omega')D_{d}
     e^{+i(\omega+\omega')\tau}
     ,
     \label{eq:DDdaggered_hatFrpx-linear}
\end{eqnarray}
\begin{eqnarray}
  &&
     \!\!\!\!\!\!\!\!\!\!\!\!\!\!\!\!\!\!
     D_{d}^{\dagger}\hat{F}_{rp(y)}(\Omega)D_{d}
     \nonumber\\
  &=&
      \frac{\hbar}{2c}
      e^{+i\theta/2}
      e^{+i\Omega\tau}
      \int
      \frac{d\omega}{2\pi}
      \sqrt{|\omega(\Omega-\omega)|}
      \left[
      \hat{D}_{c}(\omega)
      \hat{D}_{c}(\Omega-\omega)
      \right.
      \nonumber\\
  && \quad\quad\quad\quad\quad\quad\quad\quad\quad
     \left.
     +
     2
     \hat{D}_{c}(\omega)
     \left(
     \hat{D}_{v}(\Omega-\omega)
     -
     \hat{A}(\Omega-\omega)
     \right)
     \right]
     \nonumber\\
  &&
     +
     i
     \frac{\hbar}{2c^{2}}
     e^{+i\theta/2}
     \iint
     \frac{d\omega}{2\pi}
     \frac{d\omega'}{2\pi}
     \sqrt{|\omega\omega'|}
     (\omega+\omega')
     \hat{D}_{c}(\omega)
     \hat{D}_{c}(\omega')
     \nonumber\\
  && \quad\quad\quad\quad\quad\quad\quad\quad\quad
     \times
     D_{d}^{\dagger}\hat{Z}_{y}(\Omega-\omega-\omega')D_{d}
     e^{+i(\omega+\omega')\tau}
     .
     \label{eq:DDdaggered_hatFrpy-linear}
\end{eqnarray}


Through the expressions of the radiation-pressure forces
(\ref{eq:DDdaggered_hatFrpx-linear}) and
(\ref{eq:DDdaggered_hatFrpy-linear}),
Eqs.~(\ref{eq:DDdaggered-eq-of-motion-com-freq.-domain}) and
(\ref{eq:DDdaggered-eq-of-motion-dif-freq.-domain}) of the end-mirrors
are given by
\begin{eqnarray}
  &&
     \!\!\!\!\!\!\!\!\!\!\!\!\!\!\!\!\!\!
     m \Omega^{2}
     D_{d}^{\dagger}
     \hat{Z}_{com}(\Omega)
     D_{d}
     \nonumber\\
  &=&
      -
      \frac{\hbar}{c}
      e^{+i\Omega\tau}
      \cos\left(\frac{\theta}{2}\right)
      \int
      \frac{d\omega}{2\pi}
      \sqrt{|\omega(\Omega-\omega)|}
      \hat{D}_{c}(\omega)
      \hat{D}_{c}(\Omega-\omega)
  \nonumber\\
  &&
      -
     \frac{2\hbar}{c}
     e^{+i\Omega\tau}
     \int
     \frac{d\omega}{2\pi}
     \sqrt{|\omega(\Omega-\omega)|}
     \hat{D}_{c}(\omega)
     \left(
     \cos\left(\frac{\theta}{2}\right)
     \hat{D}_{v}(\Omega-\omega)
     \right.
     \nonumber\\
  && \quad\quad\quad\quad\quad\quad\quad\quad\quad\quad\quad\quad\quad\quad\quad\quad\quad
     \left.
          -
          i
          \sin\left(\frac{\theta}{2}\right)
          \hat{A}(\Omega-\omega)
        \right)
  \nonumber\\
  &&
     -
     i
     \cos\left(\frac{\theta}{2}\right)
     \frac{\hbar}{c^{2}}
     \iint
     \frac{d\omega}{2\pi}
     \frac{d\omega'}{2\pi}
     \sqrt{|\omega\omega'|}
     (\omega+\omega')
     \hat{D}_{c}(\omega)
     \hat{D}_{c}(\omega')
  \nonumber\\
  && \quad\quad\quad\quad\quad\quad\quad\quad\quad\quad\quad\quad
     \times
     D_{d}^{\dagger}\hat{Z}_{com}(\Omega-\omega-\omega')D_{d}
     e^{+i(\omega+\omega')\tau}
     \nonumber\\
  &&
     -
     \sin\left(\frac{\theta}{2}\right)
     \frac{\hbar}{c^{2}}
     \iint
     \frac{d\omega}{2\pi}
     \frac{d\omega'}{2\pi}
     \sqrt{|\omega\omega'|}
     (\omega+\omega')
     \hat{D}_{c}(\omega)
     \hat{D}_{c}(\omega')
  \nonumber\\
  && \quad\quad\quad\quad\quad\quad\quad\quad\quad\quad\quad\quad
     \times
     D_{d}^{\dagger}\hat{Z}_{diff}(\Omega-\omega-\omega')D_{d}
     e^{+i(\omega+\omega')\tau}
     ,
  \label{eq:DDdaggered-eq-of-motion-com-freq.-domain-exp-raid.press.}
\end{eqnarray}
\begin{eqnarray}
  &&
     \!\!\!\!\!\!\!\!\!\!\!\!\!\!\!\!\!\!
  m \Omega^{2}
  D_{d}^{\dagger}
  \hat{Z}_{diff}(\Omega)
  D_{d}
     \nonumber\\
  &=&
      i \sin\left(\frac{\theta}{2}\right)
      \frac{\hbar}{c}
      e^{+i\Omega\tau}
      \int
      \frac{d\omega}{2\pi}
      \sqrt{|\omega(\Omega-\omega)|}
      \hat{D}_{c}(\omega)
      \hat{D}_{c}(\Omega-\omega)
      \nonumber\\
  &&
      +
      \frac{2\hbar}{c}
      e^{+i\Omega\tau}
      \int
      \frac{d\omega}{2\pi}
      \sqrt{|\omega(\Omega-\omega)|}
     \hat{D}_{c}(\omega)
     \left(
       i \sin\left(\frac{\theta}{2}\right)
       e^{+i\theta/2}
       \hat{D}_{v}(\Omega-\omega)
     \right.
     \nonumber\\
  && \quad\quad\quad\quad\quad\quad\quad\quad\quad\quad\quad\quad\quad\quad\quad\quad\quad
     \left.
       -
       \cos\left(\frac{\theta}{2}\right)
       \hat{A}(\Omega-\omega)
     \right)
     \nonumber\\
  &&
     -
     \sin\left(\frac{\theta}{2}\right)
     \frac{\hbar}{c^{2}}
     \iint
     \frac{d\omega}{2\pi}
     \frac{d\omega'}{2\pi}
     \sqrt{|\omega\omega'|}
     (\omega+\omega')
     \hat{D}_{c}(\omega)
     \hat{D}_{c}(\omega')
  \nonumber\\
  && \quad\quad\quad\quad\quad\quad\quad\quad\quad\quad\quad\quad
     \times
     D_{d}^{\dagger}\hat{Z}_{com}(\Omega-\omega-\omega')D_{d}
     e^{+i(\omega+\omega')\tau}
     \nonumber\\
  &&
     -
     i
     \cos\left(\frac{\theta}{2}\right)
     \frac{\hbar}{c^{2}}
     \iint
     \frac{d\omega}{2\pi}
     \frac{d\omega'}{2\pi}
     \sqrt{|\omega\omega'|}
     (\omega+\omega')
     \hat{D}_{c}(\omega)
     \hat{D}_{c}(\omega')
  \nonumber\\
  && \quad\quad\quad\quad\quad\quad\quad\quad\quad\quad\quad\quad
     \times
     D_{d}^{\dagger}\hat{Z}_{diff}(\Omega-\omega-\omega')D_{d}
     e^{+i(\omega+\omega')\tau}
     \nonumber\\
  &&
     +
     \frac{1}{2} m L \Omega^{2} h(\Omega)
     .
  \label{eq:DDdaggered-eq-of-motion-dif-freq.-domain-exp-raid.press.}
\end{eqnarray}


The explicit representations of
$D_{d}^{\dagger}\hat{Z}_{com}(\Omega)D_{d}$ and
$D_{d}^{\dagger}\hat{Z}_{diff}(\Omega)D_{d}$ are given as the solutions
to
Eqs.~(\ref{eq:DDdaggered-eq-of-motion-com-freq.-domain-exp-raid.press.})
and
(\ref{eq:DDdaggered-eq-of-motion-dif-freq.-domain-exp-raid.press.}),
respectively.
Through these solutions, we can evaluate the input-output relation
(\ref{eq:linear-hatB-coherent-state-Heisenberg}) in a closed form.


\subsection{Remarks}
\label{sec:Remarks}


Here, we describe some remarks on the results of this section.


In Eq.~(\ref{eq:linear-hatB-coherent-state-Heisenberg}), we extend the
state of the photon field from the light source from the monochromatic
laser, which is described by the $\delta$-function complex amplitude for
the coherent state whose support only at the $\omega=\omega_{0}$, to
an arbitrary coherent state whose complex amplitude is described by an
arbitrary function $\alpha(\omega)$ in the frequency domain.
Due to this extension, in
Eq.~(\ref{eq:linear-hatB-coherent-state-Heisenberg}), we have to
evaluate the convolution in the frequency domain, while the
integration for this convolution is simplified due to the
$\delta$-function in the conventional Michelson gravitational-wave
detectors.
Since the complex amplitude for the coherent state from the light
source is arbitrary in
Eq.~(\ref{eq:linear-hatB-coherent-state-Heisenberg}), we do not have
central frequency $\omega_{0}$ of the complex amplitude, the
sideband picture around this central frequency, nor the approximation
$\omega_{0}\gg\Omega$.
These are due to the fact that we did not use the two-photon
formulation.


Furthermore, we introduce the phase offset $\theta$ in the
input-output relation
(\ref{eq:linear-hatB-coherent-state-Heisenberg}).
Due to this phase offset $\theta$,
Eq.~(\ref{eq:linear-hatB-coherent-state-Heisenberg}) has some effects
which are not taken into account in the input-output relation for the
conventional Michelson gravitational-wave detector.
The first one is the leakage of the classical carrier field from the
light source and the second one is the shot noise from the light
source.
These are described by the first and second terms in the right-hand
side of Eq.~(\ref{eq:linear-hatB-coherent-state-Heisenberg}), respectively.
The final one is the effect of common motion of the two end-mirrors, which
described by the second line in
Eq.~(\ref{eq:linear-hatB-coherent-state-Heisenberg}).
This common motion $D_{d}^{\dagger}\hat{Z}_{com}(\Omega)D_{d}$
together with the differential motion
$D_{d}^{\dagger}\hat{Z}_{diff}(\Omega)D_{d}$ in
Eq.~(\ref{eq:linear-hatB-coherent-state-Heisenberg}) is determined by
the equations of motion of two end-mirrors which are given by Eqs.~(\ref{eq:DDdaggered-eq-of-motion-com-freq.-domain-exp-raid.press.})
and
(\ref{eq:DDdaggered-eq-of-motion-dif-freq.-domain-exp-raid.press.}).


The equations of motion
(\ref{eq:DDdaggered-eq-of-motion-com-freq.-domain-exp-raid.press.})
and
(\ref{eq:DDdaggered-eq-of-motion-dif-freq.-domain-exp-raid.press.})
are also modified due to our extension.
For example, in the equation
(\ref{eq:DDdaggered-eq-of-motion-dif-freq.-domain-exp-raid.press.})
for the differential motion of the end-mirrors, we have to evaluate
the convolution in the first four lines in the right-hand side due to
the extension of the complex amplitude for the coherent state from the
$\delta$-function $\delta(\omega-\omega_{0})$ to an arbitrary function
$\alpha(\omega)$ in the frequency domain.
Furthermore, the third and fourth lines in the right-hand side of
Eq.~(\ref{eq:DDdaggered-eq-of-motion-dif-freq.-domain-exp-raid.press.})
appears due to this extension.
These terms do not appear in the equations of motion for the
end-mirrors in the conventional Michelson gravitational-wave detector,
and arise due to the modulation of the shape the complex amplitude by
the retarded effect due to the optical field propagation from the
central beam splitter to the end-mirrors.
Moreover, the direct effect due to the classical carrier part from the
light sources and the effect due to the shot noise from the light
source appears which affects to the differential motion of end-mirrors
through the introduction of the phase offset $\theta$.


Thus, we regard that the set of the input-output relation for the
extended Michelson interferometer
(\ref{eq:linear-hatB-coherent-state-Heisenberg}), equations
(\ref{eq:DDdaggered-eq-of-motion-com-freq.-domain-exp-raid.press.})
and
(\ref{eq:DDdaggered-eq-of-motion-dif-freq.-domain-exp-raid.press.})
of motions for the end-mirrors is the main result of this paper.


\section{Re-derivation of conventional input-output relation}
\label{sec:Re-derivation_of_conventional_input-output_relation}


In this section, we show that the derived input-output relation
(\ref{eq:linear-hatB-coherent-state-Heisenberg}) with equations
(\ref{eq:DDdaggered-eq-of-motion-com-freq.-domain-exp-raid.press.})
and
(\ref{eq:DDdaggered-eq-of-motion-dif-freq.-domain-exp-raid.press.}) of
motions yields the input-output relation for the conventional Michelson
interferometric gravitational-wave detector.


\subsection{Input-output relation}
\label{sec:Re-derivation_of_conventional_input-output_relation-1}


In the conventional Michelson gravitational-wave detector, the state
of the optical beam from the light source is in the coherent state
with the complex amplitude
\begin{eqnarray}
  \label{eq:continuous-monochromatic-carrier-field}
  \alpha(\omega) = 2\pi N \delta(\omega-\omega_{0})
  .
\end{eqnarray}
We note that $\alpha(\omega)$ is real.
The corresponding electric field with the amplitude
(\ref{eq:continuous-monochromatic-carrier-field}) of $\alpha(\omega)$
is the continuous monochromatic carrier field with the frequency
$\omega_{0}$.
The normalization factor $N$ is related to the averaged photon number
per second
\begin{eqnarray}
  \label{eq:averaged_photon_number_in_input_carrier}
  N = \sqrt{\frac{I_{0}}{\hbar\omega_{0}}},
\end{eqnarray}
where $I_{0}$ is the averaged power of the carrier field.
Through the definition (\ref{eq:hatDc-def}), the classical part
$\hat{D}_{c}(\omega)$ of the input light source is given by
\begin{eqnarray}
  \hat{D}_{c}(\omega)
  =
  2 \pi N \left\{
    \delta(\omega-\omega_{0}) \Theta(\omega)
    +
    \delta(\omega+\omega_{0}) \Theta(-\omega)
  \right\}
  .
  \label{eq:hatDc-def-delta}
\end{eqnarray}


Here, we concentrate on the mode with the frequency
$\omega_{0}\pm\Omega$.
For these sidebands, $\hat{D}_{c}(\omega)$ given by
Eq.~(\ref{eq:hatDc-def-delta}) includes terms which depend on
$2\omega_{0}\pm\Omega$.
In the time-domain, these terms includes the factor of the rapid
oscillation $e^{i2\omega_{0}t}$.
Therefore, this part can be removed in the data taking or the data
analyses processes and we ignore these terms, since we concentrate
only on the fluctuations with the frequency $\omega_{0}\pm\Omega$.
For this reason, $\hat{D}_{c}(\omega)$ with $\omega_{0}\pm\Omega$ may
be regarded as
\begin{eqnarray}
  \label{eq:upper-lower-sideband-carrier-field}
  \hat{D}_{c}(\omega_{0}\pm\Omega) = 2\pi N \delta(\Omega).
\end{eqnarray}
Through the same approximation, the input-output relations
(\ref{eq:linear-hatB-coherent-state-Heisenberg}) with
$\omega=\omega_{0}\pm\Omega$ are given by
\begin{eqnarray}
  &&
     \!\!\!\!\!\!\!\!\!\!\!\!\!\!\!\!\!\!
  e^{- 2 i(\omega_{0}\pm\Omega)\tau} D_{d}^{\dagger}\hat{B}(\omega_{0}\pm\Omega)D_{d}
     \nonumber\\
  &=&
      i \sin\left(\frac{\theta}{2}\right) \hat{D}_{c}(\omega_{0}\pm\Omega)
      +
      i \sin\left(\frac{\theta}{2}\right) \hat{D}_{v}(\omega_{0}\pm\Omega)
      +
        \cos\left(\frac{\theta}{2}\right) \hat{A}(\omega_{0}\pm\Omega)
      \nonumber\\
  &&
     +
     \frac{
     2iN\omega_{0}^{3/2}
     e^{\mp i \Omega \tau}
     }{
     c\sqrt{|\omega_{0}\pm\Omega|}
     }
     \left[
     i \sin\left(\frac{\theta}{2}\right)
     D_{d}^{\dagger}\hat{Z}_{com}(\pm\Omega)D_{d}
     \right.
     \nonumber\\
  && \quad\quad\quad\quad\quad\quad\quad\quad
     \left.
     -
     \cos\left(\frac{\theta}{2}\right)
     D_{d}^{\dagger}\hat{Z}_{diff}(\pm\Omega)D_{d}
     \right]
     ,
     \label{eq:linear-hatB-monochro-continuous-upper-lower}
\end{eqnarray}
and Eqs.~(\ref{eq:DDdaggered-eq-of-motion-com-freq.-domain-exp-raid.press.})
and
(\ref{eq:DDdaggered-eq-of-motion-dif-freq.-domain-exp-raid.press.})
are given by
\begin{eqnarray}
  &&
     \!\!\!\!\!\!\!\!\!\!\!\!\!\!\!\!\!\!
  m \Omega^{2}
  D_{d}^{\dagger}
  \hat{Z}_{com}(\Omega)
  D_{d}
     \nonumber\\
  &=&
      -
      \frac{
      \hbar
      N
      e^{+i\Omega\tau}
      \sqrt{\omega_{0}}
      }{
      c
      }
      \cos\left(\frac{\theta}{2}\right)
      \left(
      \sqrt{|\Omega-\omega_{0}|}
      \hat{D}_{c}(\Omega-\omega_{0})
      \right.
      \nonumber\\
  && \quad\quad\quad\quad\quad\quad\quad\quad\quad\quad\quad\quad
     \left.
      +
      \sqrt{|\Omega+\omega_{0}|}
      \hat{D}_{c}(\Omega+\omega_{0})
      \right)
      \nonumber\\
  &&
     -
     \frac{
     2
     \hbar
     N
     e^{+i\Omega\tau}
     \sqrt{\omega_{0}}
     }{
     c
     }
     \left\{
     \sqrt{|\Omega-\omega_{0}|}
     \left(
     \cos\left(\frac{\theta}{2}\right) \hat{D}_{v}(\Omega-\omega_{0})
      \right.
      \right.
      \nonumber\\
  && \quad\quad\quad\quad\quad\quad\quad\quad\quad\quad\quad\quad\quad\quad
     \left.
     \left.
     -
     i \sin\left(\frac{\theta}{2}\right) \hat{A}(\Omega-\omega_{0})
     \right)
     \right.
     \nonumber\\
  && \quad\quad\quad\quad\quad\quad\quad\quad
     \left.
     +
     \sqrt{|\Omega+\omega_{0}|}
     \left(
     \cos\left(\frac{\theta}{2}\right) \hat{D}_{v}(\Omega+\omega_{0})
      \right.
      \right.
      \nonumber\\
  && \quad\quad\quad\quad\quad\quad\quad\quad\quad\quad\quad\quad\quad\quad
     \left.
     \left.
     -
     i \sin\left(\frac{\theta}{2}\right) \hat{A}(\Omega+\omega_{0})
     \right)
     \right\}
     ,
     \label{eq:DDdaggered-eq-of-motion-com-freq.-domain-monochro-continuous}
\end{eqnarray}
\begin{eqnarray}
  &&
     \!\!\!\!\!\!\!\!\!\!\!\!\!\!\!\!\!\!
  m \Omega^{2}
  D_{d}^{\dagger}
  \hat{Z}_{diff}(\Omega)
  D_{d}
     \nonumber\\
  &=&
      +
      \frac{
      i
      \hbar
      N
      e^{+i\Omega\tau}
      \sqrt{\omega_{0}}
      }{
      c
      }
      \sin\left(\frac{\theta}{2}\right)
      \left\{
      \sqrt{|\Omega-\omega_{0}|}
      \hat{D}_{c}(\Omega-\omega_{0})
      \right.
      \nonumber\\
  && \quad\quad\quad\quad\quad\quad\quad\quad\quad\quad\quad\quad
     \left.
      +
      \sqrt{|\Omega+\omega_{0}|}
      \hat{D}_{c}(\Omega+\omega_{0})
      \right\}
      \nonumber\\
  &&
     +
     \frac{
     2
     \hbar
     N
     e^{+i\Omega\tau}
     \sqrt{\omega_{0}}
     }{
     c
     }
     \left\{
     \sqrt{|\Omega-\omega_{0}|}
     \left[
     i \sin\left(\frac{\theta}{2}\right) \hat{D}_{v}(\Omega-\omega_{0})
      \right.
      \right.
      \nonumber\\
  && \quad\quad\quad\quad\quad\quad\quad\quad\quad\quad\quad\quad\quad\quad
     \left.
     \left.
     -
     \cos\left(\frac{\theta}{2}\right) \hat{A}(\Omega-\omega_{0})
     \right]
     \right.
     \nonumber\\
  && \quad\quad\quad\quad\quad\quad\quad\quad\quad
     \left.
     +
     \sqrt{|\Omega+\omega_{0}|}
     \left[
     i \sin\left(\frac{\theta}{2}\right) \hat{D}_{v}(\Omega+\omega_{0})
      \right.
      \right.
      \nonumber\\
  && \quad\quad\quad\quad\quad\quad\quad\quad\quad\quad\quad\quad\quad\quad
     \left.
     \left.
     -
     \cos\left(\frac{\theta}{2}\right) \hat{A}(\Omega+\omega_{0})
     \right]
     \right\}
     \nonumber\\
  &&
     +
     \frac{1}{2}
     m L
     \Omega^{2}
     h(\Omega)
     .
     \label{eq:DDdaggered-eq-of-motion-dif-freq.-domain-monochro-continuous}
\end{eqnarray}


Here, we consider the situation where $\omega_{0}\gg\Omega$ and we
apply the approximation in which $\omega_{0}\pm\Omega$ in the
coefficients of the input-output relation are regarded as
$\omega_{0}\pm\Omega\sim\omega_{0}$.
Furthermore, we use
\begin{eqnarray}
  \label{eq:arm-tune-condition}
  \omega_{0}\tau = \omega_{0} \frac{L}{c} = 2 n \pi, \quad n\in{\mathbb N},
\end{eqnarray}
so that the anti-symmetric port is the dark port.
We also introduce the following variables
\begin{eqnarray}
  \label{eq:kappa-hSQL-defs}
  \kappa
  :=
  \frac{8 \omega_{0} I_{0}}{mc^{2}\Omega^{2}}
  , \quad
  h_{SQL}
  :=
  \sqrt{\frac{8\hbar}{m\Omega^{2}L^{2}}}
  .
\end{eqnarray}
In addition, since $\omega_{0}\gg\Omega$, we should regard
$\omega_{0}+\Omega>0$ and $\Omega-\omega_{0}<0$.
Then, the substitution of
Eqs.~(\ref{eq:DDdaggered-eq-of-motion-com-freq.-domain-monochro-continuous})
and
(\ref{eq:DDdaggered-eq-of-motion-dif-freq.-domain-monochro-continuous})
into Eq.~(\ref{eq:linear-hatB-monochro-continuous-upper-lower}) yields
the input-output relations as
\begin{eqnarray}
  D_{d}^{\dagger}\hat{b}_{\pm}D_{d}
  &=&
      \sin\left(\frac{\theta}{2}\right)
      \left(
        i
        +
        \kappa
        \cos\left(\frac{\theta}{2}\right)
      \right)
      \sqrt{\frac{I_{0}}{\hbar\omega_{0}}}
      2 \pi \delta(\Omega)
      \nonumber\\
  &&
     +
     e^{\pm 2 i\Omega\tau}
     \left[
       i \sin\left(\frac{\theta}{2}\right) \hat{d}_{\pm}
       +
       \cos\left(\frac{\theta}{2}\right) \hat{a}_{\pm}
     \right]
     \nonumber\\
  &&
     +
     \frac{\kappa e^{\pm 2 i\Omega\tau} }{2}
     \left[
       \sin\theta \left(
         \hat{d}_{\mp}^{\dagger} + \hat{d}_{\pm}
       \right)
     \right.
     \nonumber\\
  && \quad\quad\quad\quad\quad\quad
     \left.
       + i \cos\theta \left(
         \hat{a}_{\mp}^{\dagger} +  \hat{a}_{\pm}
       \right)
     \right]
     \nonumber\\
  &&
     -
     i
     \sqrt{\kappa}
     e^{\pm i\Omega\tau}
     \cos\left(\frac{\theta}{2}\right)
     \frac{h(\pm\Omega)}{h_{SQL}}
     ,
     \label{eq:linear-bear-hatb-monochro-continuous-upper-lower}
\end{eqnarray}
where $\hat{a}_{\pm}:=\hat{a}(\omega_{0}\pm\Omega)$,
$\hat{b}_{\pm}:=\hat{b}(\omega_{0}\pm\Omega)$, and
$\hat{d}_{\pm}:=\hat{d}(\omega_{0}\pm\Omega)$.
Here, we note that the carrier part in
Eq.~(\ref{eq:linear-bear-hatb-monochro-continuous-upper-lower}), which
is proportional to $\delta(\Omega)$ diverges due to the radiation-pressure
contribution $\kappa\propto\Omega^{-2}$.
Since we can predict this divergent part and can be removed by the
data taking or the data analysis processes, this carrier part is ignored.


\subsection{Two-photon formulation}
\label{sec:Re-derivation_of_conventional_input-output_relation-2}


Here, we note that the two-photon formulation is applicable in our
situation $\omega_{0}\gg\Omega$ and we introduce the operators
\begin{eqnarray}
  &&
  \hat{a}_{1} = \frac{1}{\sqrt{2}}(\hat{a}_{+}+\hat{a}_{-}^{\dagger})
  , \quad
  \hat{a}_{2} = \frac{1}{\sqrt{2}i}(\hat{a}_{+}-\hat{a}_{-}^{\dagger})
  ,
  \label{eq:hata1-hata2-def}
  \\
  &&
  \hat{b}_{1} = \frac{1}{\sqrt{2}}(\hat{b}_{+}+\hat{b}_{-}^{\dagger})
  , \quad
  \hat{b}_{2} = \frac{1}{\sqrt{2}i}(\hat{b}_{+}-\hat{b}_{-}^{\dagger})
     ,
  \label{eq:hatb1-hatb2-def}
  \\
  &&
  \hat{d}_{1} = \frac{1}{\sqrt{2}}(\hat{d}_{+}+\hat{d}_{-}^{\dagger})
  , \quad
  \hat{d}_{2} = \frac{1}{\sqrt{2}i}(\hat{d}_{+}-\hat{d}_{-}^{\dagger})
  \label{eq:hatd1-hatd2-def}
  .
\end{eqnarray}
In the two-photon formulation which treats the situation where the
carrier field is proportional to $\cos\omega_{0}t$, $\hat{a}_{1}$,
$\hat{b}_{1}$, and $\hat{d}_{1}$ are regarded as amplitude
quadratures.
On the other hand, $\hat{a}_{2}$, $\hat{b}_{2}$, and $\hat{d}_{2}$ are
regarded as phase quadratures.
Through these amplitude and phase quadratures, the input-output
relation (\ref{eq:linear-bear-hatb-monochro-continuous-upper-lower})
yields
\begin{eqnarray}
  D_{d}^{\dagger}\hat{b}_{1}D_{d}
  &=&
  \frac{1}{\sqrt{2}}
  \sin\theta
  \kappa
  \sqrt{\frac{I_{0}}{\hbar\omega_{0}}}
  2 \pi \delta(\Omega)
  \nonumber\\
  &&
  +
  e^{+ 2 i\Omega\tau}
  \left\{
    - \sin\left(\frac{\theta}{2}\right) \hat{d}_{2}
    + \cos\left(\frac{\theta}{2}\right) \hat{a}_{1}
  \right\}
  \nonumber\\
  &&
  +
  e^{+ 2 i\Omega\tau}
  \kappa
  \sin\theta
  \hat{d}_{1}
  ,
  \label{eq:linear-bear-hatb-monochro-continuous-amplitude}
  \\
  D_{d}^{\dagger}\hat{b}_{2}D_{d}
  &=&
  \sqrt{2}
  \sin\left(\frac{\theta}{2}\right)
  \sqrt{\frac{I_{0}}{\hbar\omega_{0}}}
  2 \pi \delta(\Omega)
  \nonumber\\
  &&
  +
  e^{+ 2 i\Omega\tau}
  \left\{
    \sin\left(\frac{\theta}{2}\right) \hat{d}_{1}
    +
    \cos\left(\frac{\theta}{2}\right) \hat{a}_{2}
  \right\}
  \nonumber\\
  &&
  +
  \cos\theta
  e^{+ 2 i\Omega\tau}
  \kappa
  \hat{a}_{1}
  \nonumber\\
  &&
  -
  e^{+i\Omega\tau} \cos\left(\frac{\theta}{2}\right)
  \sqrt{2\kappa} \frac{h(\Omega)}{h_{SQL}}
  .
  \label{eq:linear-bear-hatb-monochro-continuous-phase}
\end{eqnarray}


In Eq.~(\ref{eq:linear-bear-hatb-monochro-continuous-amplitude}),
the first term is the divergent classical carrier field induced by the radiation
pressure force due to the mirror motion.
The second term is the shot noise from the quantum fluctuations from
the bright port and the dark port.
The last term in
Eq.~(\ref{eq:linear-bear-hatb-monochro-continuous-amplitude}) is the
radiation pressure noise due to the mirror motion which originally
comes from the quantum fluctuation in the incident optical beam from
the bright port.
On the other hand, in
Eq.~(\ref{eq:linear-bear-hatb-monochro-continuous-phase}), the first
term is the classical carrier field which leaks from the light source
by the phase offset $\theta$.
The second term is the shot noise from the quantum fluctuation from
the bright port and the dark port.
The third line is the radiation pressure noise due to the mirror
motion which originally comes from the quantum fluctuations in the
vacuum from the dark port.
The last term is the gravitational-wave signal.


Although the classical carrier parts in
$D_{d}^{\dagger}\hat{b}_{1}D_{d}$ and
$D_{d}^{\dagger}\hat{b}_{2}D_{d}$ is completely determined in
classical sense and can be removed in the data analysis, we also
note that the classical carrier part which diverge due to the
radiation pressure force contributes only to the amplitude
quadrature $D_{d}^{\dagger}\hat{b}_{1}D_{d}$.
Therefore, as far as we observe only
$D_{d}^{\dagger}\hat{b}_{2}D_{d}$~\cite{homodyne-kouchan-preparation-comment},
the divergent term due to the
radiation pressure force cancels and we do not have to care
about this divergence.


Finally, we note that if we choose $\theta=0$,
Eqs.~(\ref{eq:linear-bear-hatb-monochro-continuous-amplitude}) and (\ref{eq:linear-bear-hatb-monochro-continuous-phase}) are reduced to
\begin{eqnarray}
  D_{d}^{\dagger}\hat{b}_{1}D_{d}
  &=&
      e^{+ 2 i\Omega\tau}
      \hat{a}_{1}
      ,
      \label{eq:linear-bear-hatb-monochro-continuous-amplitude-0offset}
  \\
  D_{d}^{\dagger}\hat{b}_{2}D_{d}
  &=&
      e^{+ 2 i\Omega\tau}
      \left(
      \hat{a}_{2}
      +
      \kappa
      \hat{a}_{1}
      \right)
      -
      e^{+i\Omega\tau}
      \sqrt{2\kappa} \frac{h(\Omega)}{h_{SQL}}
      ,
  \label{eq:linear-bear-hatb-monochro-continuous-phase-0offset}
\end{eqnarray}
respectively.
This is the conventional input-output relation for the Michelson
gravitational-wave detector~\cite{H.Miao-PhDthesis-2010}.
This means that the input-output relation
(\ref{eq:linear-bear-hatb-monochro-continuous-amplitude}) and
(\ref{eq:linear-bear-hatb-monochro-continuous-phase}) are recover the
usual input-output relation which is well-known in the
gravitational-wave community.
Furthermore, this also means that the set of the original input-output
relation (\ref{eq:linear-hatB-coherent-state-Heisenberg}) and
Eqs.~(\ref{eq:DDdaggered-eq-of-motion-com-freq.-domain-exp-raid.press.})
and
(\ref{eq:DDdaggered-eq-of-motion-dif-freq.-domain-exp-raid.press.})
of mirrors' motions are the natural extension of the conventional
input-output relation of the Michelson interferometric
gravitational-wave detector.


\section{Weak-value amplification from the extended input-output relation}
\label{sec:WVA_from_the_extended_input-output_relation}


Here, we consider the situation of the weak measurement in the
interferometer setup depicted in
Fig.~\ref{fig:kouchan-Michelson-setup-notation} from the input-output
relation (\ref{eq:linear-hatB-coherent-state-Heisenberg}) to show that
this input-output relation actually includes the weak-value amplification.
Without loss of generality, we may choose $\omega>0$ in
Eq.~(\ref{eq:linear-hatB-coherent-state-Heisenberg}):
\begin{eqnarray}
  &&
     \!\!\!\!\!\!\!\!\!\!\!\!\!\!\!\!\!\!
  e^{- 2 i \omega\tau} D_{d}^{\dagger}\hat{b}(\omega)D_{d}
     \nonumber\\
  &=&
  i \sin\left(\frac{\theta}{2}\right) \alpha(\omega)
  +
  i \sin\left(\frac{\theta}{2}\right) \hat{d}(\omega)
  +
  \cos\left(\frac{\theta}{2}\right) \hat{a}(\omega)
  \nonumber\\
  &&
  +
  \frac{2i}{c}
  \int_{-\infty}^{+\infty}
  \frac{d\Omega}{2\pi}
  e^{-i\Omega\tau}
  \sqrt{\left|\frac{\omega-\Omega}{\omega}\right|}
  (\omega-\Omega)
  \nonumber\\
  && \quad\quad\quad\quad\quad
  \times
  \left[
    i \sin\left(\frac{\theta}{2}\right)
    \hat{D}_{c}(\omega-\Omega)
    D_{d}^{\dagger}\hat{Z}_{com}(\Omega)D_{d}
  \right.
  \nonumber\\
  && \quad\quad\quad\quad\quad\quad\quad
  \left.
    -
    \cos\left(\frac{\theta}{2}\right)
    \hat{D}_{c}(\omega-\Omega)
    D_{d}^{\dagger}\hat{Z}_{diff}(\Omega)D_{d}
  \right]
  .
  \label{eq:linear-hatB-coherent-state-Heisenberg-append}
\end{eqnarray}
To discuss the weak measurement from this input-output relation, we
concentrate on the output photon number operator $\hat{n}(\omega)$ to
the photo-detector
\begin{eqnarray}
  \label{eq:output-photon-number-operator}
  \hat{n}(\omega)
  :=
  \hat{b}^{\dagger}(\omega)\hat{b}(\omega)
\end{eqnarray}
and its expectation value in the state (\ref{eq:total-state-def}) is
given by
\begin{eqnarray}
  \overline{n(\omega)}
  &:=&
       \langle\psi|\hat{n}(\omega)|\psi\rangle
       \nonumber\\
  &=&
      \langle 0|_{a}\otimes\langle 0|_{d}
      \left(
        D_{d}^{\dagger}
        \hat{b}(\omega)
        D_{d}
      \right)^{\dagger}
      \left(
        D_{d}^{\dagger}
        \hat{b}(\omega)
        D_{d}
      \right)
      |0\rangle_{a}\otimes|0\rangle_{d}
  \label{eq:output-photon-number-expectation-value}
\end{eqnarray}


Here, we consider the situation where $\hat{Z}_{com}$ and
$\hat{Z}_{diff}$ are classical, i.e., proportional to the identity
operator in the sense of quantum theory and their frequency-dependence
are negligible.
Furthermore, the complex amplitude $\alpha(\omega)$ for the coherent
state from the light source is real and has
its compact support within the frequency $\omega\in[0,\infty]$ and is
rapidly decreasing at the boundaries $\omega\rightarrow 0$ and
$\omega\rightarrow+\infty$ of this range.
This is the situation discussed by Nishizawa in
Ref.~\cite{A.Nishizawa-2015}.


Substituting
Eq.~(\ref{eq:linear-hatB-coherent-state-Heisenberg-append}) into
Eq.~(\ref{eq:output-photon-number-expectation-value}), and taking the
linear term with respect to $\hat{Z}_{com}$ and $\hat{Z}_{diff}$, we
obtain
\begin{eqnarray}
  \overline{n(\omega)}
  &=&
      \sin^{2}\left(\frac{\theta}{2}\right) \alpha^{2}(\omega)
      \nonumber\\
  &&
     -
     \sin^{2}\left(\frac{\theta}{2}\right)
     \frac{8}{2\pi c \omega^{1/2}}
     {\cal I}_{s+3/2}(\tau,\alpha)
     \alpha(\omega)
     \nonumber\\
  && \quad
     \times
     \left(
     \cos(\omega\tau)
     \hat{Z}_{com}
     +
     \cot\left(\frac{\theta}{2}\right)
     \sin(\omega\tau)
     \hat{Z}_{diff}
     \right)
     ,
     \label{eq:overline_n_omega-result}
\end{eqnarray}
where we introduce the definite integral
${\cal I}_{s+3/2}(\tau,\alpha)$ defined by
\begin{eqnarray}
  {\cal I}_{s+3/2}(\tau,\alpha)
  :=
  \int_{0}^{+\infty}
  dx x^{3/2} \sin(x\tau) \alpha(x)
  .
     \label{eq:calIs+3/2-def}
\end{eqnarray}
When $\alpha(\omega)$ is given by the Gaussian function, this definite
integral ${\cal I}_{s+3/2}(\tau,\alpha)$ does converge and expressed
using the  parabolic cylinder
function~\cite{I.S.Gradshteyn-I.M.Ryzhik-2000}.
Here, we define $\overline{n_{0}(\omega)}$ and $\delta n(\omega)$ by
\begin{eqnarray}
  \overline{n_{0}(\omega)}
  &:=&
       \sin^{2}\left(\frac{\theta}{2}\right) \alpha^{2}(\omega)
       ,
       \label{eq:overline_n0_omega-def}
  \\
  \delta n(\omega)
  &:=&
       -
       \sin^{2}\left(\frac{\theta}{2}\right)
       \frac{8}{2\pi c\sqrt{\omega}}
       {\cal I}_{s+3/2}(\tau,\alpha)
       \alpha(\omega)
      \nonumber\\
  &&
     \times
     \left(
     \cos(\omega\tau)
     \hat{Z}_{com}
     +
     \cot\left(\frac{\theta}{2}\right)
     \sin(\omega\tau)
     \hat{Z}_{diff}
     \right)
     ,
     \label{eq:delta_n_omega-def}
\end{eqnarray}
so that
\begin{eqnarray}
  \overline{n(\omega)}
  =
  \overline{n_{0}(\omega)}
  +
  \delta n(\omega)
  .
  \label{eq:overline_nomega-notation}
\end{eqnarray}


As explained in
Sec.~\ref{sec:Weak-measurement-ispired_version_of_Michelson_GW_detector},
we consider the normalized frequency distribution $f(\omega)$ of the
output photon number defined by Eq.~(\ref{eq:f-distribution-def}).
We evaluate the expectation value of the frequency $\omega$ under the
distribution function $f(\omega)$ as
\begin{eqnarray}
  \langle \omega\rangle
  &:=&
       \int_{0}^{+\infty} d\omega \omega f(\omega)
       \nonumber\\
  &\sim&
         \omega_{0}
         +
         \frac{
         \displaystyle
         \int_{0}^{+\infty} d\omega (\omega - \omega_{0}) \delta n(\omega)
         }{
         \displaystyle
         \int_{0}^{+\infty} d\omega \overline{n_{0}(\omega)}
         }
         ,
  \label{eq:expectation-of-omega-def}
\end{eqnarray}
where we denoted
\begin{eqnarray}
  \label{eq:omega0-def}
  \omega_{0}
  &:=&
       \frac{
       \displaystyle
       \int_{0}^{+\infty} d\omega \omega \overline{n_{0}(\omega)}
       }{
       \displaystyle
       \int_{0}^{+\infty} d\omega \overline{n_{0}(\omega)}
       }
       .
\end{eqnarray}
Furthermore, we introduce following definite integrals
\begin{eqnarray}
  \label{eq:calJalpha-def}
  {\cal J}(\alpha)
  &:=&
  \int_{0}^{+\infty} d\omega \alpha^{2}(\omega)
  ,
  \\
  \label{eq:calIc-1/2-def}
  {\cal I}_{c\pm 1/2}(\tau,\alpha)
  &:=&
  \int_{0}^{+\infty}
  dx x^{\pm 1/2} \cos(x\tau) \alpha(x)
  ,
  \\
  \label{eq:calIs-1/2-def}
  {\cal I}_{s\pm 1/2}(\tau,\alpha)
  &:=&
  \int_{0}^{+\infty}
  dx x^{\pm 1/2} \sin(x\tau) \alpha(x)
  .
\end{eqnarray}
When $\alpha(\omega)$ is the Gaussian function, these definite
integrals do converge.
Using these definite integrals, the expectation value
(\ref{eq:expectation-of-omega-def}) of the frequency $\omega$ under the
distribution function (\ref{eq:f-distribution-def}) is given by
\begin{eqnarray}
  \langle \omega\rangle
  -
  \omega_{0}
  &\sim&
         \hat{Z}_{com}
         \frac{8}{2\pi c {\cal J}(\alpha)}
         {\cal I}_{s+3/2}(\tau,\alpha)
         \nonumber\\
  && \quad\quad
     \times
      \left(
      \omega_{0}
      {\cal I}_{c-1/2}(\tau,\alpha)
      -
      {\cal I}_{c+1/2}(\tau,\alpha)
      \right)
      \nonumber\\
  &&
     +
     \cot\left(\frac{\theta}{2}\right)
     \hat{Z}_{diff}
     \frac{8}{2\pi c {\cal J}(\alpha)}
     {\cal I}_{s+3/2}(\tau,\alpha)
     \nonumber\\
  && \quad\quad
     \times
     \left(
     \omega_{0}
     {\cal I}_{s-1/2}(\tau,\alpha)
     -
     {\cal I}_{s+1/2}(\tau,\alpha)
     \right)
     .
     \nonumber\\
  \label{eq:expectation-of-omega-result}
\end{eqnarray}
When $\theta\ll 1$, the second term in
Eq.~(\ref{eq:expectation-of-omega-result}) is dominant, i.e.,
\begin{eqnarray}
  \langle \omega\rangle
  -
  \omega_{0}
  &\sim&
         +
         \frac{2}{\theta}
         \hat{Z}_{diff}
         \frac{8}{2\pi c {\cal J}(\alpha)}
         {\cal I}_{s+3/2}(\tau,\alpha)
         \nonumber\\
  && \quad\quad
     \times
     \left(
     \omega_{0}
     {\cal I}_{s-1/2}(\tau,\alpha)
     -
     {\cal I}_{s+1/2}(\tau,\alpha)
     \right)
     .
     \nonumber\\
  \label{eq:weak-value-amplifcation-result}
\end{eqnarray}
This is the weak-value amplification effect.


We note that there is no effect of the quantum fluctuations which
described by the quadratures $\hat{a}$ nor $\hat{d}$, but completely
determined by the amplitude $\alpha(\omega)$ of the coherent state for
the quadrature $\hat{d}$.
We also note that these arguments does not seriously depend on the
details of the real function $\alpha(\omega)$.


Thus, we have shown that our derived input-output relation
(\ref{eq:linear-hatB-coherent-state-Heisenberg}) actually includes the
situation of the weak-value amplification.


\section{Summary and discussion}
\label{sec:Summary_Discussion}


In this paper, we considered the extension of the input-output
relation for a conventional Michelson gravitational-wave detector to
compare the weak-measurement inspired gravitational-wave detector in
Ref.~\cite{A.Nishizawa-2015} with conventional one.
The main difference between these detectors is the injected optical
field, which is a continuous monochromatic laser in conventional one
and is the continuous pulse beam in the model of
Ref.~\cite{A.Nishizawa-2015}.
Therefore, we extend the conventional input-output
relation for the gravitational-wave detector to the situation where
the injected photon state is a coherent state with an arbitrary
complex amplitude $\alpha(\omega)$.
We also showed that our extended input-output relation actually
includes both situations of conventional gravitational-wave detectors
and that where the weak-value amplification occurs.
This is the main result of this paper.


Within this paper, we do not discuss quantum noise in the situation
where the weak-value amplification occurs.
However, in principle, we will be able to discuss quantum noises,
i.e., the shot noise and the radiation-pressure noise due to the
quantum fluctuations of photons in the Michelson interferometric
gravitational-wave detector, even in the situation where the
weak-value amplification occurs.
In our derivation, we regard that the Fourier transformed variables
describes the situation of the stationary continuous measurement
through the average of the many pulses.
Although this stationarity is an assumption throughout this paper, we
can discuss the time-evolution of the gravitational-wave signal
through the frequency dependence of the mirror displacement in the
extended input-output relation in Sec.~\ref{sec:Extension_of_Input-output}.
This is the difference from discussions in
Ref.~\cite{A.Nishizawa-2015} which assume that the mirror displacement
is constant in time.
In Sec.~\ref{sec:WVA_from_the_extended_input-output_relation}, we just
dare to consider the situation where the mirror displacement is constant in
time just for the comparison with Ref.~\cite{A.Nishizawa-2015}.
For this reason, we should regard that our derived input-output
relation in Sec.~\ref{sec:Extension_of_Input-output} is different from
that derived in Ref.~\cite{A.Nishizawa-2015}.


In spite of this difference from Ref.~\cite{A.Nishizawa-2015}, we
reached to the same conclusions as those in
Ref.~\cite{A.Nishizawa-2015}.
First, as discussed in
Sec.~\ref{sec:WVA_from_the_extended_input-output_relation}, the
weak-value amplification from the input-output relation
(\ref{eq:linear-hatB-coherent-state-Heisenberg}) is the effect due to
the carrier field $\alpha(\omega)$ of the coherent state from the
light source and has nothing to do with the quantum fluctuations
described by the annihilation and creation operators for photon.
Second, together with the amplification of the gravitational-wave
signal, the weak-value amplification also amplify the
radiation-pressure noise which is one of important quantum noise in
gravitational-wave detectors.
These two conclusions are not affected by the details of the analyses.
In this sense, these are robust.
In addition to the above two conclusions, from the comparison between
the input-output relations
(\ref{eq:linear-bear-hatb-monochro-continuous-upper-lower}) in
Sec.~\ref{sec:Re-derivation_of_conventional_input-output_relation} and
Eq.~(\ref{eq:linear-hatB-coherent-state-Heisenberg-append}) in
Sec.~\ref{sec:WVA_from_the_extended_input-output_relation},
we may say that a conventional Michelson gravitational-wave detector
already includes the essence of the weak-value amplification as the
reduction of the quantum noise from the light source through the
measurement at the dark port as the final conclusion.


In the situation where the weak-value amplification occurs discussed
in Sec.~\ref{sec:WVA_from_the_extended_input-output_relation}, the
unperturbed photon number, i.e., the first term in the right-hand
side of Eq.~(\ref{eq:overline_n_omega-result}), proportional to
$\sin^{2}(\theta/2)$ and there are factors $\sin^{2}(\theta/2)$ and
$\sin(\theta/2)\cos(\theta/2)$ in the coefficients of $\hat{Z}_{com}$
and $\hat{Z}_{diff}$, respectively.
Since we consider the expectation value of $\omega$ under the
conditional photon-number distribution $f(\omega)$ defined by
Eq.~(\ref{eq:f-distribution-def}), we divide the perturbed terms,
i.e.,  the second term in the right-hand side of
Eq.~(\ref{eq:overline_n_omega-result}) which are proportional to
$\hat{Z}_{com}$ or $\hat{Z}_{diff}$, by the unperturbed photon number
in Eq.~(\ref{eq:overline_n_omega-result}).
Through this process, the factor $\sin^{2}(\theta/2)$ in the
coefficient of the term including $\hat{Z}_{com}$ canceled out, but
the coefficient of the $\hat{Z}_{diff}$ becomes $\cot(\theta/2)$.
Then, if we choose $\theta\ll 1$, the term which includes $\hat{Z}_{diff}$
is dominant.
This is the weak-value amplification.
Actually, the weak value in our setup depicted in
Fig.~\ref{fig:kouchan-Michelson-setup-notation} is proportional to
$\cot(\theta/2)$ as shown in Eq.~\ref{eq:standard-weak-value}.
The important point is the fact that the weak-value amplifies
$\hat{Z}_{diff}$ which includes not only gravitational-wave signal
$h(\Omega)$ but also the radiation-pressure noise.
Therefore, we cannot improve the signal to noise ratio by the
weak-value amplification at least in the simple model in this paper as
pointed out by Nishizawa~\cite{A.Nishizawa-2015}.
Since the reduction of these noises from the dark-port is the
main target in some researches of gravitational-wave detectors,
this model is not useful for this target.


On the other hand, we compare of input-output relations
(\ref{eq:linear-bear-hatb-monochro-continuous-upper-lower}) and
(\ref{eq:linear-hatB-coherent-state-Heisenberg-append}) through the
original input-output relation
(\ref{eq:linear-hatB-monochro-continuous-upper-lower}).
The coefficients of $\hat{Z}_{com}$ and $\hat{Z}_{diff}$ in
Eq.~(\ref{eq:expectation-of-omega-result}), which yields the
weak-value amplification, are determined by the last term in the
input-output relation
Eq.~(\ref{eq:linear-hatB-coherent-state-Heisenberg-append}).
The same $(\sin(\theta/2),\cos(\theta/2))$-dependence in the
input-output relation can be seen in the original input-output
relation~(\ref{eq:linear-hatB-coherent-state-Heisenberg}).
The same dependence can also be seen in the input-output relation
(\ref{eq:linear-hatB-monochro-continuous-upper-lower}) for the
conventional gravitational-wave detector and subsequent
input-output relations
(\ref{eq:linear-bear-hatb-monochro-continuous-upper-lower}),
(\ref{eq:linear-bear-hatb-monochro-continuous-amplitude}) and
(\ref{eq:linear-bear-hatb-monochro-continuous-phase}).
Therefore, the same effect as the weak-value amplification is already
included in the conventional Michelson-interferometric
gravitational-wave detector through the photon detection at the
anti-symmetric port which are nearly dark-port.
We may say that the common motion $\hat{Z}_{com}$ and, equivalently,
the quantum fluctuations associated with the quadrature $\hat{d}$
which affect the input-output relation through $\hat{Z}_{com}$ are
negligible due to the the weak-value amplification.
This is the meaning of the above final conclusion.


Although the model discussed here is not useful for the reduction of
the quantum noise at the dark port in a gravitational-wave detector,
there are still rooms to discuss the weak-measurement inspired
gravitational-wave detector.
One of the issues to be clarified is the effect due to the optical
pulse injection from the light source instead of the monochromatic
continuous optical laser in the conventional gravitational-wave
detectors.
As the first step of this research is to examination of the
input-output relation of
(\ref{eq:linear-hatB-coherent-state-Heisenberg}) without the
approximation in which we regard that $\hat{Z}_{com}$ and
$\hat{Z}_{diff}$ are almost constant, while our arguments in
Sec.~\ref{sec:WVA_from_the_extended_input-output_relation} are based
on this approximation.
This examination will leads to discussion of the direct comparison
with the conventional Michelson-interferometric gravitational-wave
detector.
To complete this discussion, we have to examine the problem whether or
not the assumptions which are introduced when we derive the
conventional input-output relations
(\ref{eq:linear-bear-hatb-monochro-continuous-amplitude-0offset}) and
(\ref{eq:linear-bear-hatb-monochro-continuous-phase-0offset})
are also valid even in the model of Ref.~\cite{A.Nishizawa-2015}.
First, to derive these input-output relations, we concentrate on the
sidebands $\omega_{0}\pm\Omega$ as
Eq.~(\ref{eq:linear-hatB-monochro-continuous-upper-lower}).
We have to discuss the sideband picture is appropriate for the
weak-measurement or not.
As far as the discussion within the level of
Sec.~\ref{sec:WVA_from_the_extended_input-output_relation} in this
paper, we cannot apply the sideband-picture in the weak-measurement
inspired gravitational-wave detector.
Second, in the derivation of
Eq.~(\ref{eq:linear-bear-hatb-monochro-continuous-amplitude-0offset})
and (\ref{eq:linear-bear-hatb-monochro-continuous-phase-0offset}), we
ignored the high frequency modes with the frequency
$2\omega_{0}\pm\Omega$.
On the other hand, in our weak measurement model, we consider the
broad frequency distribution of the photon field in
Sec.~\ref{sec:WVA_from_the_extended_input-output_relation}.
We have to judge whether we can ignore the high frequency modes even
in the weak-measurement inspired gravitational-wave detector, or not.
Finally, we considered the situation $\omega_{0}\gg \Omega$ in the
derivation of
Eq.~(\ref{eq:linear-bear-hatb-monochro-continuous-amplitude-0offset})
and (\ref{eq:linear-bear-hatb-monochro-continuous-phase-0offset}).
This will not be appropriate in the weak-measurement inspired
gravitational-wave detector.


In addition to the above problem, the input-output relation
(\ref{eq:linear-hatB-coherent-state-Heisenberg}) might not
converge due to the response of the detector.
This will be the most important issue for the model in
Ref.~\cite{A.Nishizawa-2015}.
Roughly speaking,
$D^{\dagger}_{d}\hat{Z}_{com}(\Omega)\hat{D}_{d}$ and
$D^{\dagger}_{d}\hat{Z}_{diff}(\Omega)\hat{D}_{d}$ will have the pole
proportional to $1/\Omega^{2}$ through the equations of motion
(\ref{eq:DDdaggered-eq-of-motion-com-freq.-domain-exp-raid.press.})
and
(\ref{eq:DDdaggered-eq-of-motion-dif-freq.-domain-exp-raid.press.}).
If $\hat{D}_{c}(\omega-\Omega)$ in
Eq.~(\ref{eq:linear-hatB-coherent-state-Heisenberg}) basically given
by the Gaussian function, the integrand in
Eq.~(\ref{eq:linear-hatB-coherent-state-Heisenberg}) might diverge due
to this pole $1/\Omega^{2}$.
If this divergence is true and important in our situation, we have to
carefully discuss the physical meaning of this divergence and treat it
delicately.


We have to carefully discuss these issues for the complete
comparison between the weak-measurement inspired gravitational-wave
detector and the conventional Michelson gravitational-wave detector.
However, these issues are beyond the current scope of this paper.
Therefore, we leave this comparison with conventional
gravitational-wave detectors as one of  future works.


Even if we complete the arguments in the case where $\hat{Z}_{com}$
and $\hat{Z}_{diff}$ are not constant, we might reach to the
conclusion that the conventional Michelson-interferometric
gravitational-wave detector is more appropriate as a
gravitational-wave detector than weak-measurement inspired
gravitational-wave detectors.
However, even in this case, we will be able to discuss the effect of
the pulse-train light source.
In the experimental optics, there is a report on the ultrashort
optical pulse trains produced by the mode lock laser,  which
states that there are shot noise correlations in the frequency domain
of an ultrashort optical pulse trains and we can reduce the shot noise
using this correlation~\cite{F.Quinlan-etal.-2013}.
If we can use the same technique as this experiment, there is a
possibility to reduce the shot noise through the correlations in the
ultrashort optical pulse train produced by the mode-lock laser.
Of-course, this is no longer the weak-value amplification but this
idea comes from the point of view inspired by the weak measurements.
We will hope that our discussion in this paper will be useful when we
discuss this interesting possibility.
We also leave this interesting possibility as one of future works.


\section*{Acknowledgments}


K.N. acknowledges to Dr. Tomotada Akutsu and the other members of the
gravitational-wave project office in NAOJ for their continuous
encouragement to our research.
K.N. also appreciate Prof. Akio Hosoya and Prof. Izumi Tsutsui for
their support and continuous encouragement.





\begin{thebibliography}{00}
\bibitem{phase-offset-introduction-comment}
  In the actual gravitational-wave detectors, the similar phase offset
  is introduced in the different context.
  When the arm length $L$ is chosen so that the anti-symmetric port is
  completely dark port, the response of the interferometer to
  gravitational-wave propagation is maximum.
  However, in this maximum response, we cannot measure the phase of
  gravitational wave due to the symmetry of the interferometer.
  Since detected photon numbers are continuously monitored and the
  feedback-control technique is used so that the anti-symmetric port
  is always almost dark port with this finite phase offset.
  This feedback-control electric current is the actual data of the
  gravitational-wave observation.
  At the complete dark port, this feedback-control technique is not
  applicable.
\bibitem{beam-splitter-comment}
  In quantum mechanics,
  Eqs.~(\ref{eq:output-electric-field-junction}),
  (\ref{eq:xarm-input-field-junction}), and
  (\ref{eq:yarm-input-field-junction}) are described by the unitary
  transformation which comes from the energy conservation law.
  For an ideal beam splitter, the reflectivity of the each side of the
  beam splitter is different and the incident wave from the direction
  from the side of lower reflectivity to the side of higher
  reflectivity of the beam splitter is reflected with the $\pi$ phase
  shift or fixed end reflection.
  This is a simplest choice of the boundary condition at the beam
  splitter, though one parameter family of the boundary conditions is
  possible according to the physical property of the beam splitter.
\end{thebibliography}


\end{document}